\documentclass[12pt,onecolumn,draftclsnofoot]{IEEEtran}
\IEEEoverridecommandlockouts
\usepackage{setspace}               
\usepackage{stfloats}
\usepackage{cite}
\usepackage{amsmath,amssymb,amsfonts}
\usepackage{upgreek}
\usepackage{graphicx}
\usepackage{subfigure}
\usepackage{textcomp}
\usepackage{xcolor}
\usepackage{algorithm}
\usepackage[noend]{algpseudocode}
\usepackage{amsmath}
\usepackage[caption=false,font=normalsize,labelfont=sf,textfont=sf]{subfig}
\usepackage[linewidth=1pt]{mdframed}
\usepackage[mathscr]{eucal}
\usepackage{balance}
\usepackage{epstopdf}
\usepackage{cases}
\usepackage{xcolor}
\usepackage{bbding}
\usepackage{cleveref}
\usepackage{multicol}       
\usepackage{multirow}       
\usepackage{array}          
\usepackage{makecell}
\usepackage{booktabs}
\usepackage{flushend}
\definecolor{crimson}{RGB}{192,0,0}         
\definecolor{navy}{RGB}{47,85,151}         
\usepackage{colortbl}
\newtheorem{theorem}{\bf Theorem}
\newtheorem{corollary}{\bf Corollary}
\def\BibTeX{{\rm B\kern-.05em{\sc i\kern-.025em b}\kern-.08em
    T\kern-.1667em\lower.7ex\hbox{E}\kern-.125emX}}

\begin{document}
\title{Double-Layer Power Control for Mobile Cell-Free XL-MIMO with Multi-Agent Reinforcement Learning}
\author{{Ziheng~Liu, Jiayi~Zhang,~\IEEEmembership{Senior Member,~IEEE}, Zhilong~Liu,~\IEEEmembership{Graduate Student Member,~IEEE}, Huahua~Xiao, and Bo~Ai,~\IEEEmembership{Fellow,~IEEE}}
\thanks{Part of this article has been accepted at IEEE INFOCOM 2023 \cite{[2]}.} \thanks{Ziheng Liu, Jiayi Zhang, Zhilong Liu and Bo Ai are with the School of Electronic and Information Engineering and also with the Frontiers Science Center for Smart High-Speed Railway System, Beijing Jiaotong University, Beijing 100044, China (e-mail: \{23111013, zhangjiayi, zhilongliu, boai\}@bjtu.edu.cn).} \thanks{Huahua Xiao is with ZTE Corporation, State Key Laboratory of Mobile Network and Mobile Multimedia Technology, Shenzhen 518055, China (e-mail: xiao.huahua@zte.com.cn).}
}
\maketitle
\vspace{-1.75cm}
\begin{abstract}
Cell-free (CF) extremely large-scale multiple-input multiple-output (XL-MIMO) is regarded as a promising technology for enabling future wireless communication systems. Significant attention has been generated by its considerable advantages in augmenting degrees of freedom. In this paper, we first investigate a CF XL-MIMO system with base stations equipped with XL-MIMO panels under a dynamic environment. Then, we propose an innovative multi-agent reinforcement learning (MARL)-based power control algorithm that incorporates predictive management and distributed optimization architecture, which provides a dynamic strategy for addressing high-dimension signal processing problems. Specifically, we compare various MARL-based algorithms, which shows that the proposed MARL-based algorithm effectively strikes a balance between spectral efficiency (SE) performance and convergence time. Moreover, we consider a double-layer power control architecture based on the large-scale fading coefficients between antennas to suppress interference within dynamic systems.
Compared to the single-layer architecture, the results obtained unveil that the proposed double-layer architecture has a nearly 24\% SE performance improvement, especially with massive antennas and smaller antenna spacing.
\end{abstract}

\IEEEpeerreviewmaketitle
\vspace{-0.3cm}
\section{Introduction}
The next-generation wireless communication systems, such as the sixth generation (6G), are expected to satisfy the increasing demand for communication quality, e.g., ultra-low access latency, massive connections, and low-cost construction.
The commercialization of massive multiple-input multiple-output (mMIMO) technology has a significant impact on the rapid development of wireless networks. However, the conventional mMIMO technology cannot fully meet the stringent requirements of 6G application scenarios. Emerging technologies including cell-free (CF) mMIMO and extremely large-scale MIMO (XL-MIMO), which break the capacity limitations of conventional mMIMO, hold great promise in addressing the aforementioned challenges.

As a high-profile technology, the innovative CF mMIMO is effective in solving increased network throughput and achieving low-latency transmission by deploying a large number of geographically distributed access points (APs), compared with cellular mMIMO \cite{[10],[28]}. Similarly, the promising XL-MIMO technology inherits the architecture of conventional cellular mMIMO with the revolutionary shift in base stations (BSs). However, the communication regions vary from the far-field to the near-field due to the massive antenna deployment \cite{[11],[14]}. Moreover, thanks to the enormous spatial multiplexing and beamforming gain, these advanced technologies play a significant role in achieving a higher spectral efficiency (SE), higher energy efficiency, and reliable massive connections.

Compared with conventional CF mMIMO \cite{[26],[20],[25]}, the novel XL-MIMO deploys numerous antennas in a compact space. As such, many hardware designs have been investigated with different structures and terminologies, i.e., \itshape{large intelligent surfaces}\upshape{,} \itshape{continuous aperture MIMO}\upshape{,} \itshape{holographic MIMO}\upshape{, and} \itshape{extremely large antenna array} \upshape \cite{[11]}, for the effective realization of XL-MIMO systems.
In addition to the difference in hardware structure, the XL-MIMO for 6G not only means a sharp increase in the number of antennas but also results in a fundamental change in the electromagnetic (EM) characteristics that can be adopted to improve communication performance \cite{[1],[12],[13]}.
Furthermore, due to the unique physical architecture of XL-MIMO, near-field propagation tends to dominate, rendering the commonly adopted uniform plane wave (UPW) models no longer valid \cite{[12],[13]}. Recently, the mainstream studies on XL-MIMO systems have shifted the focus from exploiting the far-field characteristics to concentrating on the near-field ones \cite{[16],[15],[17],[18],[37]}.
For instance, the authors in \cite{[12],[13]} comprehensively reviewed the existing XL-MIMO hardware designs and discussed the unique challenges of XL-MIMO. Moreover, the authors in \cite{[12]} proposed two cases for the hybrid propagation channel modeling and the computations of effective degrees of freedom for practical scenarios, which lay the foundation for subsequent channel modeling and performance optimization. Furthermore, the authors in \cite{[16],[15]} proposed novel Fourier plane-wave stochastic scalar channel models for the single-BS single-user (UE) scenario and single-BS multi-UE scenario, respectively, which fully capture the essence of EM propagation in the near-field communication.
\vspace{-0.3cm}
\subsection{Related Work}
To achieve abundant gains and optimize the system performance, designing adaptive power control algorithms is crucial, which necessitates the application of advanced optimization methods \cite{[19],[21],[27],[31]}. On the one hand, conventional signal processing-based power control methods have been well studied in the past decades for achieving a higher SE performance at the expense of high computational complexity.
Unfortunately, these conventional methods limit the practical implementation of mMIMO systems \cite{[22]}. On the other hand, model-free machine learning-based power control methods can significantly reduce the required computational complexity while approaching the same performance as the conventional methods \cite{[29]}. However, most of the prior model-free machine learning-based studies focus on supervised learning, which is impractical since the prior optimal output data obtained in large-scale MIMO systems is challenging.
In contrast, multi-agent reinforcement learning (MARL) is a promising technique for solving high-dimensional computation challenges, which has been adopted in numerous application scenarios, e.g., sensor networks, autonomous driving, game playing, and robotics \cite{[5],[23],[33],[35]}. In particular, MARL concentrates on optimizing the goal-oriented agent strategy and learning directly from the interaction with the environment to improve the overall learning performance.
Based on the popular centralized training and decentralized execution (CTDE) network architecture, MARL approaches the optimal joint strategy. Many efficient algorithms have been derived, such as the multi-agent deep deterministic policy gradient (MADDPG), which have been successfully applied to the CF mMIMO systems to address intractable problems over recent years. For example, the pilot assignment with the MADDPG was solved to mitigate the pilot contamination, effectively reducing the computational complexity \cite{[9]}. In addition, the authors in \cite{[24]} solved a joint communication and computing resource allocation problem with the MADDPG-based algorithm to minimize energy consumption.
\vspace{-0.3cm}
\subsection{Motivations and Contributions}
After several MARL approaches have been proposed for handling straightforward multi-agent tasks, researchers have shifted their attention to large-scale multi-agent scenarios. However, real-time information interaction in real large-scale scenarios is challenging due to the comparatively high computational complexity of centralized learning in MADDPG, while fully decentralized learning cannot always guarantee convergence. Therefore, to address the above challenges, the authors in \cite{[8]} proposed a novel paradigm with the combination of fuzzy logic and MARL, by which the training amount of the MARL network is greatly reduced while the coupling of agents can be implicitly captured. Although the application of fuzzy logic can improve the implementation of mMIMO systems by reducing the computational complexity, the mapping of entities to fuzzy agents can result in unstable convergence and a slow convergence rate.
Recently, strategies to address the aforementioned challenges from the perspective of optimizing the convergence rate have been considered \cite{[6],[7]}. Instead of reducing the training amount of the MARL network to avoid affecting the convergence effect, the authors in \cite{[6]} utilized a decoupling architecture that includes global and local critic networks to accelerate the convergence rate.
Similarly, the authors in \cite{[7]} leveraged the prioritized experience selected mechanism to improve the network architecture, which optimizes the convergence rate by extracting larger loss experiences and discarding smaller loss experiences during the training phase.
Moreover, the movement of receivers is widespread in practical mMIMO systems. The major analysis based on the assumption of static scenarios is not conducive to optimizing the system performance. Therefore, it is necessary to investigate user mobility under dynamic scenarios to improve the network architecture.
Additionally, it is noteworthy that the existing uplink power control methods overlook the optimization strategy between the antennas. The interference between antennas is relative to the multipath effect, which is detrimental to the system performance improvement. Therefore, designing an effective power control scheme to allocate appropriate power to each antenna to minimize interference is crucial.

Motivated by the aforementioned observations, this paper begins with the basic schema of CF XL-MIMO systems. Specifically, from the perspective of the EM field, the increase of antennas in CF XL-MIMO is a superficial phenomenon, and the main difference with CF mMIMO is that it pushes the EM operating region from the far-field region to the near-field region, and its analysis method has changed significantly, from the original planar wavefront-based method to the spherical wavefront-based one \cite{[36]}. To strive for the potential uplink SE performance of CF XL-MIMO systems, we introduce a novel double-layer MARL-based power control method. The major contributions of this paper are given as follows:
\vspace{-0.05cm}
\begin{itemize}
\item We first investigate a CF XL-MIMO system considering user mobility and predictive management over the near-field communication domain. Then, we derive the achievable uplink SE expression and novel closed-form with maximum ratio (MR) combining.

\item We introduce a MARL-based power control method, namely MIMO-MADDPG, which combines the decoupling architecture and the prioritized experience selected mechanism. The results demonstrate that our proposed method achieves a faster convergence rate while having a performance approaching the conventional methods.

\item We propose a double-layer architecture that considers the large-scale fading (LSF) coefficients between antennas to allocate optimal power for each antenna, which is more effective in achieving excellent SE performance.
\end{itemize}

Compared to the conference version [1], which only considers static scenarios, this paper has extended it to dynamic scenarios and added a novel MARL-based power control scheme from the perspective of optimizing convergence rate. Furthermore, based on the unique near-field large-scale fading coefficients between antennas, we extend the single-layer architecture to a double-layer architecture.
\newcounter{mytempeqncnt_1}

The rest of this paper is organized as follows. In Section \uppercase\expandafter{\romannumeral2}, we consider a near-field channel model and derive a closed-form SE expression with MR combining. Then, Section \uppercase\expandafter{\romannumeral3} introduces the uplink power control problem with user mobility. Next, in Section \uppercase\expandafter{\romannumeral4}, we propose the MIMO-MADDPG method, which combines a variety of distributed optimization networks. More important, we propose a double-layer architecture based on LSF coefficients between antennas. In Section \uppercase\expandafter{\romannumeral5}, numerical results and performance analysis are provided. Finally, the major conclusions and future directions are drawn in Section \uppercase\expandafter{\romannumeral6}.

\emph{\textbf{{    Notation}}}: The boldface lowercase letters $\bf{x}$ and boldface uppercase letters $\bf{X}$ denote the column vectors and matrices, respectively. The subscripts $\left(\cdot\right)$\textsuperscript{\emph{$\ast$}}, $\left(\cdot\right)$\textsuperscript{\emph{T}} and $\left(\cdot\right)$\textsuperscript{\emph{\textrm{H}}} represent conjugate, transpose, and conjugate transpose, respectively. $\mathbb{E\{\cdot\}}$, ${\text{tr}\{\cdot\}}$, and $\triangleq$ represent the expectation operator, the trace operator, and the definitions, respectively.
The Kronecker products and the element-wise products are denoted by $\otimes$ and $\odot$, respectively.
$\text{diag}(\mathbf{A}_1,\ldots,\mathbf{A}_n)$ denotes a block-diagonal matrix. $\text{vec}(\mathbf{A})$ denote the column vector formed by stacking the columns of $\mathbf{A}$. $\mathbb{R}$ denotes the set of real numbers. $|\cdot|$ and $\|\cdot\|$ are the determinant of a matrix and the Euclidean norm, respectively. Finally, $\mathbf{x}\sim{{\cal N}_\mathbb{C}}\left({0},{\bf{R}}\right)$ represents the circularly symmetric complex Gaussian distribution vector with zero mean and correlation matrix ${{\bf{R}}}$.
\begin{figure}[t]
\centering
    \includegraphics[scale=0.15]{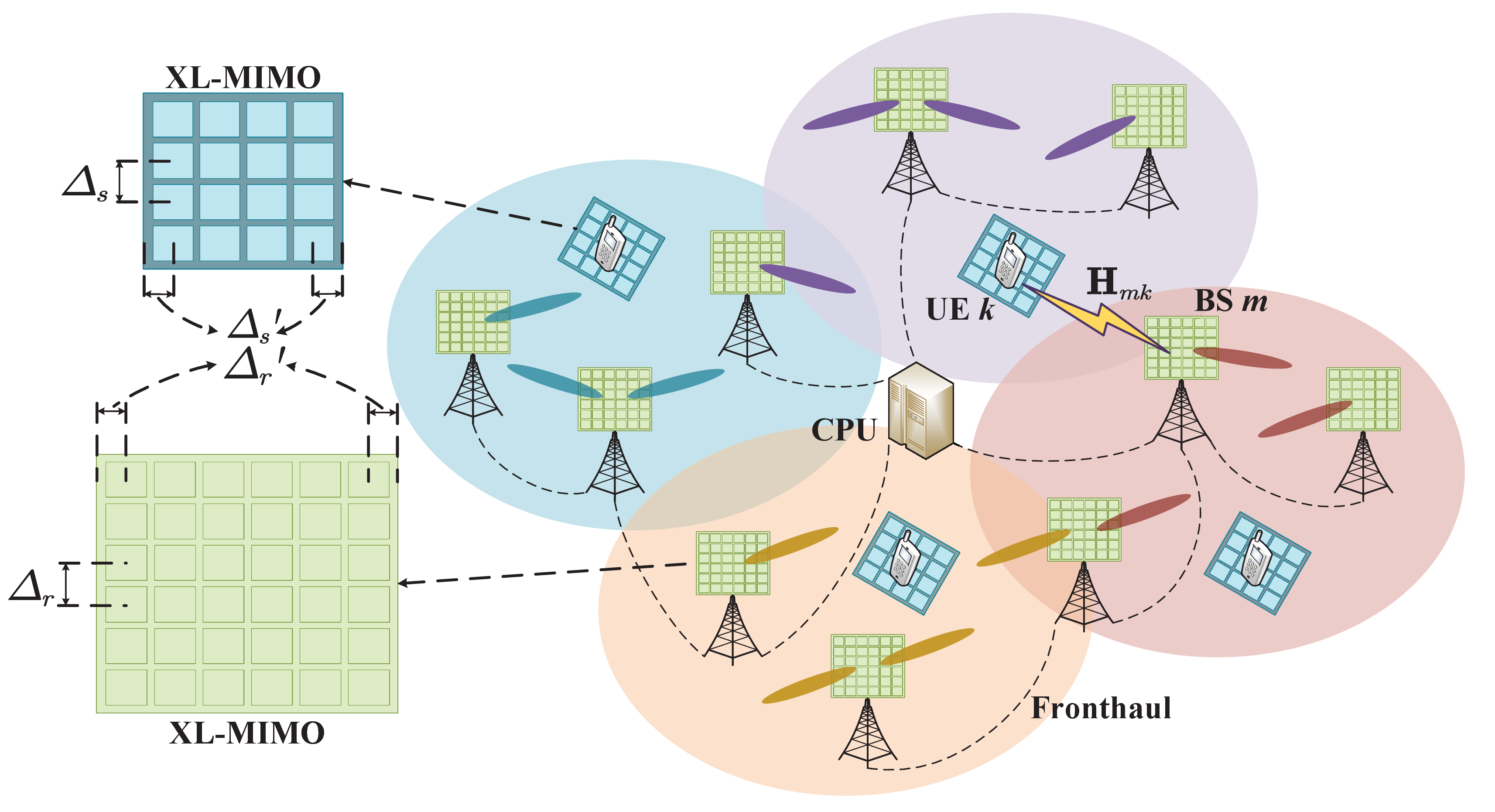}
    \caption{Illustration of a CF XL-MIMO system.
    \label{fig1}}
\end{figure}

\section{System Model}
In this paper, we investigate the uplink SE performance of a CF XL-MIMO system consisting of $M$ BSs and $K$ UEs. All BSs are connected to a central processing unit (CPU) via perfect fronthaul links \cite{[30]}, as shown in Fig. 1.
Considering the advances in antenna technology in recent years, making it possible to integrate multiple antennas into smaller and more compact devices, and the design of large user end devices including smart factories and smart transportation, we can assume that each BS and UE consists of a planar extremely large-scale surface (XL-surface), and each BS is equipped with $N_r = {N_{V_r}}{N_{H_r}}$ patch antennas, where $N_{V_r}$ and $N_{H_r}$ denote the number of antennas in the vertical and horizontal direction. We assume that all planar XL-surfaces are parallel and that the horizontal and vertical patch antenna spacing $\Delta_r$ is less than half of the carrier wavelength $\lambda$ at each BS. Hence, the horizontal and vertical length of each planar XL-surface can be denoted by $L_{r,x}=(N_{H_r}-1)\Delta_r+2\Delta_r'$ and $L_{r,y}=(N_{V_r}-1)\Delta_r+2\Delta_r'$, respectively, where $\Delta_r'$ is the edge spacing of the patch antenna.
In general, the edge spacing $\Delta_r'$ can be simplified as half the antenna spacing $\Delta_r / 2$. Therefore, the horizontal and vertical length of each planar XL-surface can be modeled as $L_{r,x}=N_{H_r}\Delta_r$ and $L_{r,y}=N_{V_r}\Delta_r$, respectively.
Additionally, the antennas at each BS are indexed row-by-row by $n_r \in [1,N_r]$, thus the location of the $n_r$-th antenna at BS $m$ with respect to the origin can be expressed in three-dimension form as $\mathbf{r}_m^{(n_r)} = [r_{m,x}^{(n_r)},r_{m,y}^{(n_r)},r_{m,z}^{(n_r)}]^T, n_r = [1,\ldots,N_r]$.

Then the received signals can be denoted as
$\!\mathbf{a}_{r}(\mathbf{k},\mathbf{r})=[\mathbf{a}_{r,1}(\mathbf{k}_1,\mathbf{r}_1), \ldots, \mathbf{a}_{r,M}(\mathbf{k}_M,\mathbf{r}_M)]\!$ with the receive signal
$\mathbf{a}_{r,m}(\mathbf{k}_m,\mathbf{r}_m)=[e^{j\mathbf{k}_{m}(\varphi_m,\theta_m)^T{\mathbf{r}_m^{(1)}}}, \ldots, e^{j\mathbf{k}_{m}(\varphi_m,\theta_m)^T{\mathbf{r}_{m}^{(N_r)}}}]^T$ at BS $m$,
where $\mathbf{k}_{m}(\varphi_m,\theta_m)=k[\cos(\theta_m)\cos(\varphi_m),$ $\cos(\theta_m)\sin(\varphi_m),\sin(\theta_m)] \in \mathbb{R}^3$
is the receive wave vector with the wavenumber $k=2\pi/\lambda$, and $\theta_m$ and $\varphi_m$ are the receive elevation angle and the receive azimuth angle at BS $m$, respectively, $\forall m \in \{1, \ldots, M\}$.

Similarly, each UE is equipped with $N_s = {N_{V_s}}{N_{H_s}}$ patch antennas, where the patch antenna spacing, horizontal length, and vertical length are $\Delta_s$, $L_{s,x}=(N_{H_s}-1)\Delta_s+2\Delta_s'=N_{H_s}\Delta_s$, and $L_{s,y}=(N_{V_s}-1)\Delta_s+2\Delta_s'=N_{V_s}\Delta_s$, respectively, with the edge spacing $\Delta_s'=\Delta_s/2$. The antennas are indexed row-by-row by $n_s \in [1,N_s]$, and the location of the $n_s$-th antenna at UE $k$ with respect to the origin can be defined as $\mathbf{s}_k^{(n_s)} = [s_{k,x}^{(n_s)},s_{k,y}^{(n_s)},s_{k,z}^{(n_s)}]^T, n_s = [1,\ldots,N_s]$.

Besides, the transmit signals can be denoted as $\mathbf{a}_{s}(\boldsymbol{\kappa},\mathbf{s})=[\mathbf{a}_{s,1}(\boldsymbol{\kappa}_1,\mathbf{s}_1), \ldots,\mathbf{a}_{s,K}(\boldsymbol{\kappa}_K,\mathbf{s}_K)]$, and the transmit signal is
$\mathbf{a}_{s,k}(\boldsymbol{\kappa}_k,\mathbf{s}_k)=[e^{j\boldsymbol{\kappa}_{k}(\varphi_k,\theta_k)^T{\mathbf{s}_k^{(1)}}}, \ldots, e^{j\boldsymbol{\kappa}_{k}(\varphi_k,\theta_k)^T{\mathbf{s}_k^{(N_s)}}}]^T$ at UE $k$,
where $\boldsymbol{\kappa}_{k}(\varphi_k,\theta_k)=k[\cos(\theta_k)\cos(\varphi_k),$
$\cos(\theta_k)\sin(\varphi_k),\sin(\theta_k)] \in \mathbb{R}^3$ is the transmit wave vector with the elevation angle $\theta_k$ and the azimuth angle $\varphi_k$ at UE $k$, $\forall k \in \{1, \ldots, K\}$.
\subsection{Single-BS Multi-UE Channel Model}
Based on the single-BS multi-UE channel model proposed by the authors in [13], it extends the individual user channel modeling from the previous subsection to the multi-user case, and it is assumed that different users are independently distributed in space. To fully characterize EM channel, each BS and UE is equipped with $N_r \geqslant \frac{4}{\lambda^2}L_{r,x}L_{r,y}$ and $N_s \geqslant \frac{4}{\lambda^2}L_{s,x}L_{s,y}$ patch antennas, respectively. Then, the channel of $k$-th UE in matrix form $\mathbf{H}^{(k)} \in \mathbb{C}^{N_r\times{N_s}}$ can be denoted as
\begin{equation}
\setcounter{equation}{1}
\begin{split}
\mathbf{H}^{(k)}=\sqrt{{N_r}{N_s}}\sum_{(\ell_x,\ell_y) \in \varepsilon_r}\sum_{(m_x,m_y) \in \varepsilon_s}H_a^{(k)}(\ell_x,\ell_y,m_x,m_y)\mathbf{a}_{r}(\ell_x,\ell_y,\mathbf{r})\mathbf{a}_{s,k}(m_x,m_y,\mathbf{s}^{(k)}).
\end{split}
\end{equation}

Note that the sparsity of the channel in the wavenumber domain is only non-zero for finite elements. Therefore, the spatial domain channel $\mathbf{H}^{(k)}$ in (1) can be approximated by the finite sampling points in the lattice ellipse, which can be defined as $\varepsilon_r=\{(\ell_x,\ell_y)\in\mathbb{Z}^2:(\ell_x\lambda/L_{r,x})^2+(\ell_y\lambda/L_{r,y})^2\leqslant1\}$ and $\varepsilon_s=\{(m_x,m_y)\in\mathbb{Z}^2:(m_x\lambda/L_{s,x})^2+(m_y\lambda/L_{s,y})^2\leqslant1\}$. And the receive wave vector $\mathbf{a}_{r}(\mathbf{k},\mathbf{r})$ and the transmit wave vector $\mathbf{a}_{s,k}(\mathbf{\kappa}_k,\mathbf{s}_k)$ can be denoted as
\begin{equation}
\setcounter{equation}{2}
\begin{split}
\left \{
\begin{array}{ll}
\mathbf{a}_{r}^{n_r}(\ell_x,\ell_y,\mathbf{r}) = \frac{1}{N_r}e^{-j\left(\frac{2\pi}{L_{r,x}}\ell_x r_{x}^{(n_r)} + \frac{2\pi}{L_{r,y}}\ell_y r_{y}^{(n_r)} + \sqrt{(\frac{2\pi}{\lambda})^2-\ell_x^2 -\ell_y^2} r_{z}^{(n_r)}\right)}, n_r = [1,\ldots,N_r],\\
\mathbf{a}_{s,k}^{n_s}(m_x,m_y,\mathbf{s}^{(k)}) = \frac{1}{N_s}e^{-j\left(\frac{2\pi}{L_{s,x}}m_x s_{k,x}^{(n_s)} + \frac{2\pi}{L_{s,y}}m_y s_{k,y}^{(n_s)} + \sqrt{(\frac{2\pi}{\lambda})^2-m_x^2 -m_y^2} s_{k,z}^{(n_s)}\right)}, n_s = [1,\ldots,N_s].
\end{array}
\right.
\end{split}
\end{equation}
where $H_a^{(k)}(\ell_x,\ell_y,m_x,m_y)$ is the Fourier coefficient with $\sigma_{(k)}^2(\ell_x,\ell_y,m_x,m_y)$, satisfying
\begin{equation}
\setcounter{equation}{3}
H_a^{(k)}(\ell_x,\ell_y,m_x,m_y) \sim \mathcal{N}_\mathbb{C}(0,\sigma_{(k)}^2(\ell_x,\ell_y,m_x,m_y)).
\label{eq3}
\end{equation}

Following similar steps in [13], $\mathbf{U}_{r} \in \mathbb{C}^{N_r\times n_r}$ and $\mathbf{U}_{s,k}\in \mathbb{C}^{N_s\times n_s}$ denote the matrices collecting the variances of $n_r$-th and $n_s$-th sampling points in $(\ell_x,\ell_y,\mathbf{r})$ and $(m_x,m_y,\mathbf{s}_k)$, respectively. Then the channel matrix of $k$-th UE for the single-BS multi-UE scenario can be approximated as
\begin{equation}
\setcounter{equation}{4}
\mathbf{H}^{(k)}=\mathbf{U}_{r}\mathbf{H}_a^{(k)}(\mathbf{U}_{s,k})^H=\mathbf{U}_{r}(\mathbf{Q}_{k}\odot \mathbf{W})(\mathbf{U}_{s,k})^H,
\label{eq4}
\end{equation}
where
$\mathbf{H}_a^{(k)} = \mathbf{Q}_{k}\odot \mathbf{W} \in \mathbb{C}^{n_r\times n_s}$ collects $\sqrt{N_rN_s}$ $H_a^{(k)}(\ell_x,\ell_y,m_x,m_y)$ for all $n_r\cdot n_s$ sampling points with $\mathbf{W} \sim \mathcal{N}_\mathbb{C}(0,\mathbf{I}_{n_rn_s})$. And $\mathbf{Q}_{k}=(\mathbf{v}_{r}\mathbf{1}_{n_s}^T)\odot(\mathbf{1}_{n_r}\mathbf{v}_{s,k}^T)$ where $\mathbf{v}_{r} \in \mathbb{R}^{n_r\times1}$ and $\mathbf{v}_{s,k} \in \mathbb{R}^{n_s\times1}$ collect $\sqrt{N_r}\sigma_{r}(\ell_x,\ell_y)$ and $\sqrt{N_s}\sigma_{s,k}(m_x,m_y)$, respectively.
\vspace{-0.1cm}
\subsection{Multi-BS Multi-UE Channel Model}
Furthermore, based on the single-BS multi-UE model derived above, the corresponding small-scale fading (SSF) coefficient $\mathbf{H}_{mk} \in \mathbb{C}^{N_r\times{N_s}}$ for the multi-BS multi-UE scenario can be denoted as
\begin{equation}
\setcounter{equation}{5}
\begin{split}
\mathbf{H}_{mk}&=\sqrt{{N_r}{N_s}}\sum_{(\ell_x,\ell_y) \in \varepsilon_r}\sum_{(m_x,m_y) \in \varepsilon_s}H_a^{(mk)}(\ell_x,\ell_y,m_x,m_y)\mathbf{a}_{r,m}(\ell_x,\ell_y,\mathbf{r}_m)\mathbf{a}_{s,k}(m_x,m_y,\mathbf{s}_k),
\end{split}
\end{equation}
where the Fourier coefficient
\begin{equation}
\setcounter{equation}{6}
H_a^{(mk)}(\ell_x,\ell_y,m_x,m_y) \sim \mathcal{N}_\mathbb{C}(0,\sigma_{mk}^2(\ell_x,\ell_y,m_x,m_y)),
\label{eq6}
\end{equation}
and the wave vector can be denoted as
\begin{equation}
\setcounter{equation}{7}
\begin{split}
\left \{
\begin{array}{ll}
\mathbf{a}_{r,m}^{n_r}(\ell_x,\ell_y,\mathbf{r}_m) = \frac{1}{N_r}e^{-j\left(\frac{2\pi}{L_{r,x}}\ell_x r_{m,x}^{(n_r)} + \frac{2\pi}{L_{r,y}}\ell_y r_{m,y}^{(n_r)} + \sqrt{(\frac{2\pi}{\lambda})^2-\ell_x^2 -\ell_y^2} r_{m,z}^{(n_r)}\right)}, n_r = [1,\ldots,N_r],\\
\mathbf{a}_{s,k}^{n_s}(m_x,m_y,\mathbf{s}_k) = \frac{1}{N_s}e^{-j\left(\frac{2\pi}{L_{s,x}}m_x s_{k,x}^{(n_s)} + \frac{2\pi}{L_{s,y}}m_y s_{k,y}^{(n_s)} + \sqrt{(\frac{2\pi}{\lambda})^2-m_x^2 -m_y^2} s_{k,z}^{(n_s)}\right)}, n_s = [1,\ldots,N_s].
\end{array}
\right.
\end{split}
\label{eq7}
\end{equation}

Similarly, the SSF model in (5) can be written as $\mathbf{H}_{mk}=\mathbf{U}_{r,m}(\mathbf{H}_a^{(mk)})(\mathbf{U}_{s,k})^H$, where $\mathbf{H}_a^{(mk)} = \mathbf{Q}_{mk}\odot \mathbf{W}_{mk} \in \mathbb{C}^{n_r\times n_s}$ collects $\sqrt{N_rN_s}H_a^{(mk)}(\ell_x,\ell_y,m_x,m_y)$ for all $n_r\cdot n_s$ sampling points with $\mathbf{W}_{mk} \sim \mathcal{N}_\mathbb{C}(0,\mathbf{I}_{n_rn_s})$. And $\mathbf{Q}_{mk}=(\mathbf{v}_{r,m}\mathbf{1}_{n_s}^T)\odot(\mathbf{1}_{n_r}\mathbf{v}_{s,k}^T)$ where $\mathbf{v}_{r,m} \in \mathbb{R}^{n_r\times1}$ and $\mathbf{v}_{s,k} \in \mathbb{R}^{n_s\times1}$ collect $\sqrt{N_r}\sigma_{r,m}(\ell_x,\ell_y)$ and $\sqrt{N_s}\sigma_{s,k}(m_x,m_y)$, respectively.

More important, based on the SSF coefficient $\mathbf{H}_{mk}=\mathbf{U}_{r,m}(\mathbf{Q}_{mk}\odot \mathbf{W}_{mk})(\mathbf{U}_{s,k})^H$, we can derive the SSF channel model $\mathbf{h}_{mk}= \text{vec}(\mathbf{H}_{mk}) \sim \mathcal{N}_\mathbb{C}(0,\mathbf{R}_{mk})\in \mathbb{C}^{N_rN_s}$ with the corresponding full correlation matrix $\mathbf{R}_{mk} = \mathbb{E}\{\text{vec}(\mathbf{H}_{mk})\text{vec}(\mathbf{H}_{mk})^H\}$ given by
\begin{equation}
\setcounter{equation}{8}
\begin{split}
\mathbf{R}_{mk}={\bigg({\mathbf{U}_{s,k}}^\ast\otimes\mathbf{U}_{r,m}\bigg)}{\bigg(\text{diag}{\Big(\mathbf{v}_{s,k}\odot\mathbf{v}_{s,k}\Big)}\otimes\text{diag}{\Big(\mathbf{v}_{r,m}\odot\mathbf{v}_{r,m}\Big)}\bigg)}{\bigg({\mathbf{U}_{s,k}}^T\otimes{\mathbf{U}_{r,m}}^H\bigg)}.
\end{split}
\label{eq8}
\end{equation}

Moreover, as the coordinate of the $n_r$-th antenna of BS $m$ and the $n_s$-th antenna of UE $k$ are
$\mathbf{r}_{m}^{(n_r)}=[r_{m,x}^{(n_r)},r_{m,y}^{(n_r)},r_{m,z}^{(n_r)}]^T$ and $\mathbf{s}_{k}^{(n_s)}=[s_{k,x}^{(n_s)},s_{k,y}^{(n_s)},s_{k,y}^{(n_s)}]^T$, respectively. Then the LSF coefficient can be denoted as
\begin{equation}
\setcounter{equation}{9}
\mathbf{B}_{mk}^{(n_r,n_s)}=\sqrt{G_tF(\vartheta_{mk}^{(n_r,n_s)})}\lambda/({4\pi d_{mk}^{(n_r,n_s)}}) \in \mathbb{C}^{N_r\times{N_s}},
\label{eq9}
\end{equation}
where $d_{mk}^{(n_r,n_s)}=\|\mathbf{r}_{m}^{(n_r)}-\mathbf{s}_{k}^{(n_s)}\|$ denotes the distance, and $\vartheta_{mk}^{(n_r,n_s)}$ denotes the corresponding angle \cite{[3]}. In addition, $G_t$ represents the antenna gain and $F(\vartheta)$ is the normalized power radiation pattern that satisfies
$\int_{0}^{\pi/2}G_tF(\vartheta)\sin\vartheta d\vartheta =1$. Based on the above derivation of LSF and SSF model, the corresponding channel model $\mathbf{G}_{mk}$ can be denoted as $\mathbf{G}_{mk} = \mathbf{B}_{mk} \odot \mathbf{H}_{mk} \in \mathbb{C}^{N_r\times{N_s}}$ \cite{[17]}.

Generally, the distance $d_{mk}$ between BS $m$ and UE $k$ is larger than the array aperture $D$. For example, for a $16\times 16$-element planar XL-surface working at 30 GHz, $d_{mk}$ is very likely to be larger than $D=16\times 0.5 \times 10^{-2}=0.08$ meters.
With $d_{mk} > D$, we can assume $\mathbf{B}_{mk}^{(1,1)}\approx\cdots\approx\mathbf{B}_{mk}^{(N_r,N_s)}\approx\beta_{mk}=\sqrt{G_tF(\vartheta_{mk})}\lambda/4\pi d_{mk}$ based on the Fresnel approximation \cite{[32]}.
Therefore, the corresponding channel model can be represented as $\mathbf{G}_{mk,s}=\beta_{mk}\mathbf{H}_{mk} \in \mathbb{C}^{N_r\times{N_s}}$.
\subsection{Uplink Data Transmission}
In CF XL-MIMO systems, all UEs simultaneously transmit their data symbols to BSs for uplink data transmission.
The transmitted signal $\mathbf{s}_k=[s_{k,1},\ldots,s_{k,N_s}]^T \in \mathbb{C}^{N_s}$ from UE $k$ can be
constructed as $\mathbf{s}_k=\mathbf{P}_k\mathbf{x}_k$, where $\mathbf{x}_k \sim \mathcal{N}_\mathbb{C}(0,\mathbf{I}_{N_s})$ is the transmitted data symbol and $\mathbf{P}_k \in \mathbb{C}^{{N_s}\times{N_s}}$ is antenna transmit power matrix during the uplink data transmission phase, satisfying the power constraint as $\text{tr}(\mathbf{P}_k\mathbf{P}_k^H) \leqslant p_k$ with $p_k$ being the upper limit of transmission power of UE $k$, respectively.
The received signal $\mathbf{y}_m \in \mathbb{C}^{N_r}$ at BS $m$ is
\begin{equation}
\setcounter{equation}{10}
\mathbf{y}_m=\sum_{k=1}^{K}{\mathbf{G}_{mk}}\mathbf{s}_k+\mathbf{n}_m=\sum_{k=1}^{K}{\mathbf{G}_{mk}}\mathbf{P}_k\mathbf{x}_k+\mathbf{n}_m,
\label{eq10}
\end{equation}
where $\mathbf{n}_m \sim \mathcal{N}_\mathbb{C}(0,\sigma^2\mathbf{I}_{N_r})$ is the independent receiver noise with the noise power $\sigma^2$.

Let $\mathbf{V}_{mk} \in \mathbb{C}^{{N_r}\times{N_s}}$ denote the combining matrix designed by BS $m$ for UE $k$. Then, the local estimation $\check{\mathbf{x}}_{mk}$ of the transmitted symbol $\mathbf{x}_k$ for UE $k$ at BS $m$ is
\begin{equation}
\setcounter{equation}{11}
\begin{aligned}
\check{\mathbf{x}}_{mk}={\mathbf{V}_{mk}^H}\mathbf{G}_{mk}\mathbf{P}_k\mathbf{x}_k+\sum_{l=1,l{\neq}k}^{K}\mathbf{V}_{mk}^H\mathbf{G}_{ml}\mathbf{P}_l\mathbf{x}_l+{\mathbf{V}_{mk}^H}\mathbf{n}_m.
\label{eq11}
\end{aligned}
\end{equation}

We notice that the common large-scale fading decoding method necessitates a significant amount of LSF knowledge. This knowledge grows quadratically with $M$, $K$, $N_r$, and $N_s$, which can become very large in CF XL-MIMO systems \cite{[4]}. In practical scenarios, the large number of LSF parameters need to be jointly estimated by the BSs and sent to the CPU, which may not be feasible, especially if the channel statistics vary with time. To simplify the signal processing process, the CPU can alternatively weight the local processed signal $\check{\mathbf{x}}_{mk}$ by averaging the observations from all $M$ BSs to obtain the final signal as
\begin{equation}
\setcounter{equation}{12}
\begin{split}
\hat{\mathbf{x}}_{k}=\frac{1}{M}\sum_{m=1}^{M}\check{\mathbf{x}}_{mk}=\frac{1}{M}\Big(\sum_{m=1}^{M}{\mathbf{V}_{mk}^H}\mathbf{G}_{mk}\mathbf{P}_k\mathbf{x}_k+\sum_{m=1}^{M}
\sum_{l=1,l{\neq}k}^{K}{\mathbf{V}_{mk}^H}\mathbf{G}_{ml}\mathbf{P}_l\mathbf{x}_l+\sum_{m=1}^{M}{\mathbf{V}_{mk}^H}\mathbf{n}_m\Big).
\end{split}
\end{equation}

Based on (12), we can derive the uplink achievable SE as the following corollary.
\begin{corollary}
\emph{An achievable SE of UE k in the CF XL-MIMO is}
\begin{equation}
\setcounter{equation}{13}
\begin{aligned}
\text{SE}_k = \log_{2}{\left|\mathbf{I}_{N_s}+\mathbf{E}_k^H\mathbf{\Psi}_k^{-1}\mathbf{E}_k\right|},
\label{eq8}
\end{aligned}
\end{equation}
\emph{where} $\mathbf{E}_k \triangleq \sum_{m=1}^{M}\mathbb{E}\{{\mathbf{V}_{mk}^H}\mathbf{G}_{mk}\}\mathbf{P}_{k}$
\emph{,} $\mathbf{\Psi}_k \triangleq$ $\sum_{l=1}^{K}\sum_{m=1}^{M}\sum_{m'=1}^{M}\mathbb{E}\{{\mathbf{V}_{mk}^H}\mathbf{G}_{ml}\mathbf{\bar{P}}_{l}\mathbf{G}_{m'l}^H\mathbf{V}_{m'k}\}-\mathbf{E}_k\mathbf{E}_k^H$
+ $\sum_{m=1}^{M}\mathbb{E}\{{\mathbf{V}_{mk}^H}\mathbf{n}_m\mathbf{n}_m^H\mathbf{V}_{mk}\}$ \emph{, and} $\mathbf{\bar{P}}_{l} \triangleq \mathbf{P}_l\mathbf{P}_l^H$.
\end{corollary}

Note that the equation (13) is applicable along with any combining scheme, such as MR combining with $\mathbf{V}_{mk}=\mathbf{G}_{mk}$, and local minimum mean-squared error combining.
Considering that MR combining does not necessitate any matrix inversion, which results in lower computational complexity \cite{[4]}. Therefore, it is more suitable for implementation in CF XL-MIMO systems, and we can obtain the closed-form SE expression with MR combining as the following theorem.
\begin{theorem}
\emph{For MR combining, we can derive the closed-form SE expression as}
\begin{equation}
\setcounter{equation}{14}
\begin{aligned}
\text{SE}_{k,c} = \log_{2}{\left|\mathbf{I}_{N_s}+\mathbf{E}_{k,c}^H\mathbf{\Psi}_{k,c}^{-1}\mathbf{E}_{k,c}\right|},
\label{eq9}
\end{aligned}
\end{equation}
\emph{where} $\mathbf{E}_{k,c} = \sum_{m=1}^{M}\mathbf{Z}_{mk}\mathbf{P}_{k}$
\emph{and} $\mathbf{\Psi}_{k,c} =\sum_{l=1}^{K}\mathbf{T}_{kl}-\mathbf{E}_k\mathbf{E}_k^H$
+ $\sigma^2\sum_{m=1}^{M}\mathbf{Z}_{mk}$.
\end{theorem}
\begin{IEEEproof}
The proof follows from the similar approach as [36] and is therefore omitted.
\end{IEEEproof}

\section{Basic Schemes of MARL-Based Method}
Considering that unreasonable power allocation leads to serious inter-user interference, which can ultimately hinder performance improvement. To optimize the system performance, designing a reasonable power control is particularly important in CF XL-MIMO systems.
\vspace{-0.2cm}
\subsection{Uplink Power Control}
In CF XL-MIMO systems, the number of antennas deployed at each BS is usually large, far exceeding the limit of $50$ antennas set in channel hardening, which will cause channel variation to decrease as more antennas are added, in a sense that the normalized instantaneous channel gain converges to the deterministic average channel gain. Therefore, we optimize the transmit power from UEs according to the LSF coefficients under the power constraint condition $\text{tr}(\mathbf{P}_k\mathbf{P}_k^H) \leqslant p_k, \forall k \in [1,K]$. As previously mentioned, the uplink power control optimization problem can be modeled as follows
\begin{equation}
\setcounter{equation}{15}
\begin{aligned}
&\quad\max_{\{\mathbf{P}_k:\forall k\}} \sum_{k=1}^{K} \text{SE}_{k,c}=\sum_{k=1}^{K}\log_{2}{\left|\mathbf{I}_{N_s}+\mathbf{E}_{k,c}^H\mathbf{\Psi}_{k,c}^{-1}\mathbf{E}_{k,c}\right|}\\
&\qquad \mbox{s.t.} \qquad \quad \text{tr}(\mathbf{P}_k\mathbf{P}_k^H) \leqslant p_k, \quad k = 1, \ldots, K.
\label{eq10}
\end{aligned}
\end{equation}

It is obvious that the power control problem in (15) is non-convex, and the computational complexity of the conventional methods is prohibitively high, rendering the original solutions incompatible with CF XL-MIMO systems. Therefore, in the following subsection, we introduce a novel MARL-based method that overcomes the aforementioned challenges.
\vspace{-0.2cm}
\subsection{Markov Decision Process Model}
Recently, RL algorithms have been applied to optimize resource allocation in mobile systems, and many efficient algorithms have been derived such as DDPG, Twin Delayed Deep Deterministic Policy Gradient (TD3), and Proximal Policy Optimization (PPO) \cite{[34],[38]}. However, in CF XL-MIMO systems, considering that TD3 and PPO involve more complex optimization processes, this can lead to instability or non-convergence of the learning process. Motivated by this trend, we can map conventional multi-agent systems into CF XL-MIMO systems in this paper based on the DDPG algorithm. Therefore, we propose a distributed MARL-based method to solve the uplink power control problem, since it enables multiple agents to complete complex tasks through collaborative decision-making in high-dimensional dynamic scenarios.

More important, the MARL is near the optimal joint policy based on the most efficient training mechanism of the Centralized Training and Decentralized Execution (CTDE). CTDE is based on the Actor-Critic architecture and all agents are composed of an \itshape Actor \upshape network for action assignment and a \itshape Critic \upshape network for policy update. This approach effectively solves the problems of non-stationary and experience playback failure in multi-agent environments.

In a multi-agent environment, all agents are treated as entities that interact directly with the environment, and a complete Markov decision process (MDP) model is the premise for the convergence of designed algorithms.
By strictly following the basic Markov chain, our designed algorithm can converge with sufficient interaction, allowing each agent to form a complete MDP model with the environment and complete the mapping of their policies to actions.

Moreover, we describe all UEs as agents with a MARL tuple
$<\mathcal{S}, \mathcal{A}, \mathcal{R}, \mathcal{P}, \gamma>$, where state space $\mathcal{S} = [s_{0},\ldots,s_{t},\ldots]$ with the observed state ${s}_{t}=[{s}_{1,t}, \ldots, {s}_{K,t}]$ and action space $\mathcal{A} = [a_{0},\ldots,a_{t},\ldots]$ with the assigned action ${a}_{t}=[{a}_{1,t}, \ldots, {a}_{K,t}]$ at $t$ time slot, depending on the LSF coefficients and the uplink power allocation coefficients, respectively.
The reward functions are $\mathcal{R} = [r_{0},\ldots,r_{t},\ldots]$ with the reward ${r}_{t}=[{r}_{1,t}, \ldots, {r}_{K,t}]$ at $t$ time slot. Furthermore, $\mathcal{P}:(\mathcal{S},\mathcal{A})\rightarrow\mathcal{S}$ is the state transition function, and $\gamma$ is the discounted factor. Then, we can model our objective in a multi-agent environment by referring to (16) as $\mathbb{E}[\mathcal{R}] = \sum_{t=t_0}^{T}\gamma^{t-t_0}r_t=\sum_{t=t_0}^{T}\gamma^{t-t_0}\sum_{k=1}^{K} \text{SE}_{k,c}^{(t)}$, where $t_0$ and $T$ are the current time and the terminal time, respectively.
\vspace{-0.2cm}
\subsection{User Mobility}
The current studies on solving mMIMO problems using MARL methods \cite{[9],[24]} focus on static scenarios, ignoring user mobility. This causes the MARL tuple $<\mathcal{S}, \mathcal{A}, \mathcal{R}, \mathcal{P}, \gamma>$ to degenerate into $<\mathcal{S}, \mathcal{A}, \mathcal{R}, \gamma>$, which completely overlooks the impact of the state transition function $\mathcal{P}$. Therefore, we combine user mobility with MARL-based power control methods and achieve predictive management in dynamic scenarios. Furthermore, the observed state $s_t$ at $t$ time slot is determined by the simplified inter-user LSF coefficients.

Then, we denote $a_t^{p}$, $a_t^{d}$, and $a_t^{\theta}$ as the power factor, distance factor, and angle factor, respectively. In conventional static scenarios, the allocated action $a_t$ belongs to the one-dimensional variable generated based on the observed state $s_t$ at time $t$, namely the power coefficient $a_t = a_t^{p}$.
However, in a dynamic scenario, the \itshape Actor \upshape network not only allocates power coefficient $a_t^{p}$ but also provides corresponding feedback for moving step $a_t^{d}$ and moving angle $a_t^{\theta}$, resulting in the action belonging to a three-dimensional variable $a_t = [a_t^{p};a_t^{d};a_t^{\theta}]$.

Therefore, the agents will update the next state $s_{t+1}$ after allocating actions $a_t = (a_t^{p},a_t^{d},a_t^{\theta})$ in each episode. And the location of UE $k$ at $t$ time slot can be modeled as
$l_{k,t}=[l_{k,t,x},l_{k,t,y},l_{k,t,z}]=[u_{k,t,x} + s_{k,x}^{(0)},u_{k,t,y}+s_{k,y}^{(0)},u_{k,t,z}+s_{k,z}^{(0)}]$. Besides, based on the assigned moving step $a_{k,t}^{d}$ and moving angle $a_{k,t}^{\theta}$ at UE $k$, the location of UE $k$ at $t+1$ time slot can be updated 
\begin{equation}
\setcounter{equation}{16}
\begin{split}
l_{k,t+1}=\bigg[u_{k,t,x} + s_{k,x}^{(0)} + a_{k,t}^{d}\cos\Big(a_{k,t}^{\theta}\Big),u_{k,t,y}+s_{k,y}^{(0)} + a_{k,t}^{d}\sin\Big(a_{k,t}^{\theta}\Big),u_{k,t,z}+s_{k,z}^{(0)}\bigg].
\end{split}
\end{equation}
where $u_{k,t,x}$, $u_{k,t,y}$, and $u_{k,t,z}$ denote the 3D locations of UE $k$ at $t$ time slot. $s_{k,x}^{(0)}$, $s_{k,y}^{(0)}$ and $s_{k,z}^{(0)}$ denote the 3D locations of the origin at UE $k$.

Then, combined with the simplified inter-user LSF coefficients and the location of UE $k$, we can leverage the state transition function $\mathcal{P}$ to obtain the state $s_{k,t+1}$ at $t+1$ time slot, the transition relationship can be modeled as $\mathcal{P}:(s_{k,t},a_{k,t})\rightarrow s_{k,t+1},k=[1,\ldots,K]$.
\vspace{-0.2cm}
\subsection{Predictive Management}
Additionally, it is important to notice that the dynamic environment of CF XL-MIMO systems in this paper differs from the common multi-agent environment. In each episode of the training stage, a common multi-agent system will evaluate whether it achieves the final target before taking the next action.
In this paper, we break the limit of the incentive ceiling and stop point for agents, which leads to a dynamic movement for better incentive results in each episode.

According to the analysis above, there is no theoretical optimal stop point in the MARL system. The established network cannot guarantee that all agents always move in the better direction and may force some agents to move from the better to the worse point.
Therefore, it is necessary to add predictive management to restrict the movement of all agents and set a reasonable threshold value for all agents to move or stop.
Besides, we denote the deceleration moving threshold at the advantage and the acceleration moving threshold at the disadvantage as $r_{g}$ and $r_{b}$. Herein, the moving step restriction relationship of the agent can be defined as
\begin{equation}
\setcounter{equation}{17}
\begin{split}
m_{k,t}= \left \{
\begin{array}{ll}
    \alpha a_{k,t}^d,   &\quad \sum_{k=1}^{K} r_{k,t} \geqslant r_{g},\\
    a_{k,t}^d,          & r_{b} < \sum_{k=1}^{K} r_{k,t} < r_{g},\\
    \beta a_{k,t}^d ,   &\quad \sum_{k=1}^{K} r_{k,t} \leqslant r_{b},
\end{array}
\right.
\label{eq12}
\end{split}
\end{equation}
where $m_{k,t}$ is the generated movement step by prediction management. $\alpha$ and $\beta$ are restriction factors of stopping movement and accelerating movement, satisfying $0 \leqslant \alpha \leqslant 1$ and $\beta > 1$, respectively. Besides, the new location of UE $k$ at $t+1$ time slot $l_{k,t+1}^{predict}$ can be updated to 
\begin{equation}
\setcounter{equation}{18}
\begin{split}
l_{k,t+1}^{predict}=\bigg[u_{k,t,x} + s_{k,x}^{(0)} + m_{k,t}\cos\Big(a_{k,t}^{\theta}\Big),u_{k,t,y}+s_{k,y}^{(0)} + m_{k,t}\sin\Big(a_{k,t}^{\theta}\Big),u_{k,t,z}+s_{k,z}^{(0)}\bigg].
\end{split}
\end{equation}
\begin{figure*}[t]
    \centering
    \includegraphics[width=5.5in]{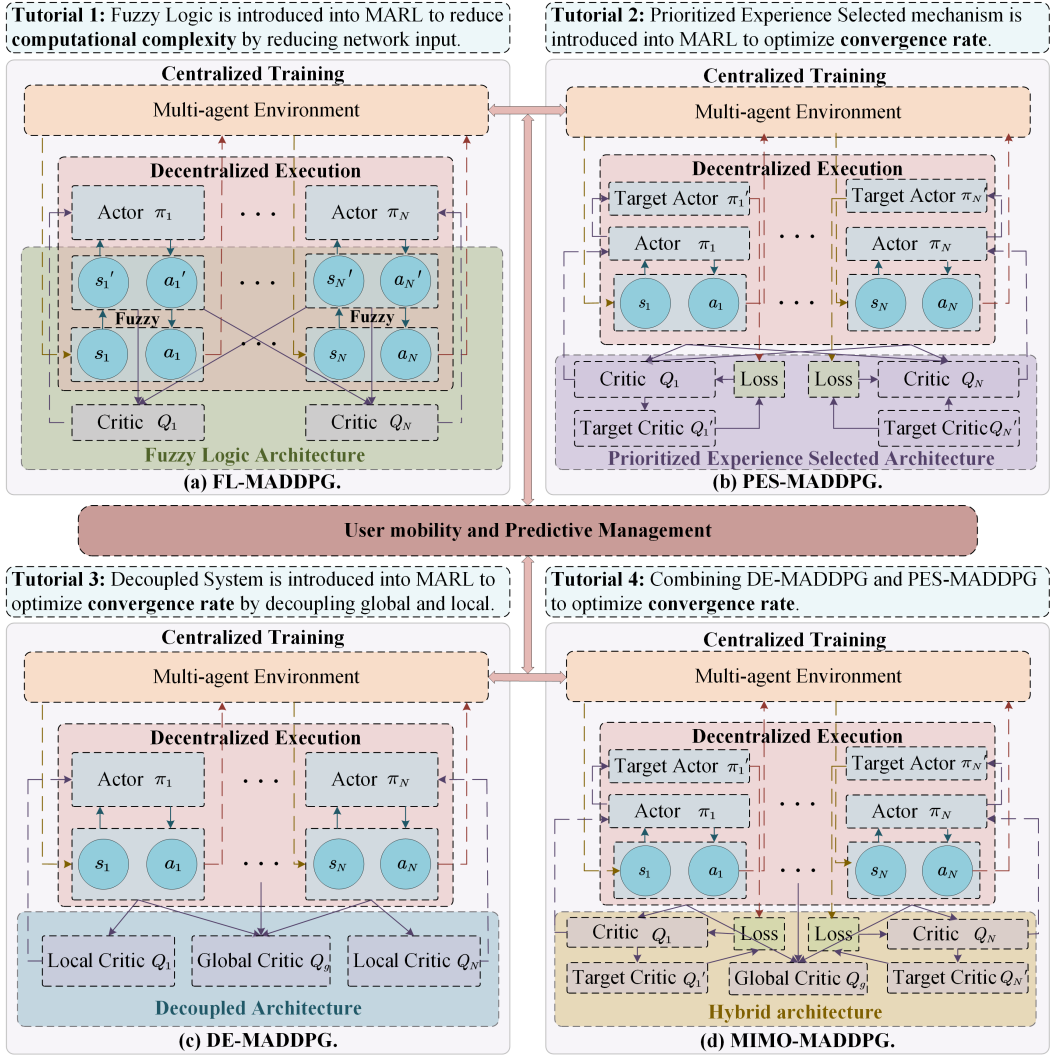}
    \caption{Illustration of four signal architectures. The FL-MADDPG architecture utilizes fuzzy logic to optimize computational complexity, while the remaining three architectures PES-MADDPG, DE-MADDPG, and MIMO-MADDPG utilize the global critic network, prioritized experience selected mechanism, and hybrid architecture to optimize convergence rate, respectively.}
    \label{fig2}
\end{figure*}
\section{Proposed MIMO-MADDPG for Maximizing SE}
In this section, we propose a novel paradigm for large-scale MARL called MIMO-MADDPG to accelerate the convergence rate, which combines the global critical network in the DE-MADDPG, the mechanism of prioritized experience selected in the PES-MADDPG, and predictive management for mobile CF XL-MIMO systems.
\begin{table*}[t]
  \centering
  \fontsize{9}{11.2}\selectfont
  \caption{Comparison of computational complexity.}
  \label{Paper_comparison}
    \begin{tabular}{ !{\vrule width1.2pt}  m{2.5 cm}<{\centering} !{\vrule width1.2pt}   m{4.5 cm}<{\centering} !{\vrule width1.2pt} m{4 cm}<{\centering} !{\vrule width1.2pt} m{4 cm}<{\centering} !{\vrule width1.2pt}}
    \Xhline{1.2pt}
        \rowcolor{gray!30} \bf Methods  &  \bf MARL-based neural network &  \bf Additional system network &  \bf Reward function \cr
    \Xhline{1.2pt}
        \bf MADDPG   & $\mathcal{O}(\sum_{a=1}^{A}MK^{3}Q_a^{2}d_a+\sum_{c=1}^{C}(MK^{2}+Kd_a)Q_c^{2})$ & --------- & $\mathcal{O}(M^{2}K^{2}+MKN_rN_s)$ \cr\hline
        \bf FL-MADDPG   & $\mathcal{O}(\sum_{a=1}^{A}MKF^{2}Q_a^{2}d_a+\sum_{c=1}^{C}(MF^{2}+Fd_a)Q_c^{2})$ & $\mathcal{O}(MKFd_a)$ & $\mathcal{O}(M^{2}K^{2}+MKN_rN_s)$ \cr\hline
        \bf PES-MADDPG & $\mathcal{O}(\sum_{a=1}^{A}MK^{3}Q_a^{2}d_a+\sum_{c=1}^{C}(MK^{2}+Kd_a)Q_c^{2})$ & $\mathcal{O}(K^{2}N_{batch}^{\mathcal{B}})$ & $\mathcal{O}(M^{2}K^{2}+MKN_rN_s)$ \cr\hline
        \bf DE-MADDPG  & $\mathcal{O}(\sum_{a=1}^{A}MK^{3}Q_a^{2}d_a+\sum_{c=1}^{C}(MK^{2}+Kd_a)Q_c^{2})$ & $\mathcal{O}(\sum_{c=1}^{C}(MK^{2}+Kd_a)Q_c^{2})$ & $\mathcal{O}(M^{2}K^{2}+MKN_rN_s)$ \cr\hline
        \bf MIMO-MADDPG  & $\mathcal{O}(\sum_{a=1}^{A}MK^{3}Q_a^{2}d_a+\sum_{c=1}^{C}(MK^{2}+Kd_a)Q_c^{2})$ & $\mathcal{O}(K^{2}N_{batch}^{\mathcal{B}}+\sum_{c=1}^{C}(MK^{2}+Kd_a)Q_c^{2})$ & $\mathcal{O}(M^{2}K^{2}+MKN_rN_s)$\cr\hline
    \Xhline{1.2pt}
    \end{tabular}
  \vspace{0cm}
\end{table*}
\subsection{Single-Layer Power Control}
For the conventional MARL-based methods, due to their high computational complexity and slow convergence rate, the action allocation and policy update of numerous agents make the designed algorithms unable to be implemented in actual large-scale scenarios. Therefore, the trend is to develop an algorithm with a faster convergence rate, real-time information exchange capability, and stability.

In this subsection, to comply with the corresponding development trend, we introduce a variety of novel distributed optimization architectures to improve the conventional network architecture and propose an extended version of the conventional MADDPG algorithm, namely MIMO-MADDPG, as shown in Fig. 2. Taking the MIMO-MADDPG algorithm as an example, we combine the advantages of DE-MADDPG and PES-MADDPG for optimizing the convergence rate. The network architecture includes a global centralized \itshape Critic \upshape network shared amongst all agents and $K$ local decentralized \itshape Critic \upshape networks leveraged by all agents independently. Essentially, the MIMO-MADDPG algorithm remains a partial architecture of CTDE to accelerate the convergence rate by adopting the decoupling architecture and the prioritized experience selected mechanism.
In addition, to mitigate the impact of outdated decisions in the MARL environment, on the one hand, we deploy a large number of antennas at each BS so that the normalized instantaneous channel gain will converge to a determined average channel gain, and on the other hand, we introduce a slower learning rate and predictive management architecture in the original network architecture to ensure that the transmission power is always optimized for instantaneous system conditions.

During the training phase, each agent extracts its own experience from the experience pool to train the \itshape Actor \upshape and \itshape Critic \upshape networks. The prioritized experience selected mechanism is combined to improve the quality of the extracted experience.
In the experience content accessed by the experience pool, in addition to the current state $s_t$, the allocated action $a_t$, the reward $r_t$, and the next state $s_{t+1}$, data such as the target evaluation network loss $loss_t=[loss_{1,t},\ldots,loss_{K,t}]$, the experience extraction training times $n_t=[n_{1,t},\ldots,n_{K,t}]$, and the priority of the current experience $Pr_t=[Pr_{1,t},\ldots,Pr_{K,t}]$ are also saved. This indicates that the replay buffer meets $\mathcal{D}=<s_t,a_t,r_t,s_{t+1},loss_t,n_t,Pr_t>$.
The greater the loss $loss_t$, the greater the difference between the evaluation value and the current value of the target network under the experience.

The priority of the experience $Pr_t$ is determined solely based on the target evaluation network loss $loss_t$, which is the only indicator used to measure the importance of experience. The relationship between them can be modeled as
\vspace{-0.1cm}
\begin{equation}
\setcounter{equation}{19}
\begin{aligned}
Pr_{k,t}=\frac{loss_{k,t}}{\sum_{k=1}^{K}loss_{k,t}}.
\label{eq14}
\end{aligned}
\end{equation}

Moreover, it is noteworthy that the loss $loss_t$ varies greatly among different experiences, and it may be difficult to extract and train partial experience depending on low $Pr_t$ solely. Therefore, we denote $rank(\varepsilon)$ as dimensionless sort quantity, where $rank(loss_{k,t})$ is the position of $loss_{k,t}$ in the ascending order of sequence $loss_t$.
Although only taking $loss_t$ as the basis for the evaluation experience and discarding the experience with smaller $loss_t$ can accelerate the training process, part of the experience may never be extracted during training, which may lead to over-fitting of the neural network or falling into local optimization. Therefore, while measuring the priority $Pr_t$, we need to comprehensively consider the loss $loss_t$ and the numbers of experience extraction training $n_t$. The corresponding priority $Pr_t$ can be updated to
\vspace{-0.1cm}
\begin{equation}
\setcounter{equation}{20}
\begin{aligned}
Pr_{k,t}=\frac{pr_{k,t}^\mu}{\sum_{k=1}^{K}pr_{k,t}^\mu}+\nu,
\label{eq15}
\end{aligned}
\end{equation}
where $pr_{k,t}=rank(rank(loss_{k,t}))+rank_{reverse}(n_{k,t})$ with $rank_{reverse}(n_{k,t})$ is the position of the extraction number $n_{k,t}$ in the descending sort $n_{t}$. Furthermore, $\mu$ is the amplification number of priority, and $\nu \in (0,1)$ is the offset of probability to prevent a lower probability of experience being selected due to a smaller $pr_{k,t}$.

Furthermore, it should be noted that before sampling for training, all experiences must have calculated the target evaluation network loss $loss_t$. Then, we use equation (20) to calculate the priority $Pr_t$ of each extracted experience from $\mathcal{D}$ and put them into the experience extraction pool $\mathcal{B}$ according to their priority, until $\mathcal{B}$ reaches the specified size of $N_{batch}^{\mathcal{B}}$.

As the processing flow of the MIMO-MADDPG is shown in Fig. 2(d), all agents are deployed at the UEs. Consequently, all UEs independently complete the allocated action $a_t$ based on local information, while the CPU uniformly completes the policy update based on global information.
In essence, the MIMO-MADDPG algorithm based on MADDPG still adheres to the \itshape actor-critic \upshape approach. Its main idea is to combine the MADDPG (the current global \itshape evaluation Critic \upshape network ${\theta_{Q_\pi}^g}$ with an additional global \itshape target Critic \upshape network ${\theta_{Q_{\pi'}}^g}$) with $K$ single-agent DDPG (the current local \itshape evaluation Actor \upshape network ${\theta_\pi}$ and local \itshape evaluation Critic \upshape network ${\theta_{Q_\pi}}$ with an additional local \itshape target Actor \upshape network ${\theta_{\pi'}}$ and local \itshape target Critic \upshape network ${\theta_{Q_{\pi'}}}$).
Then, the policy gradient of the local actor for $\pi_i$ can be modeled as 
\begin{equation}
\setcounter{equation}{21}
\begin{aligned}
\nabla_{\theta_{\pi_i}}J(\theta_{\pi_i})&=
\underbrace{\mathbb{E}_{s_t,a_t\sim \mathcal{B}}\Big[\nabla_{\theta_{\pi_i}}\pi_i({a}_{i,t}|{s}_{i,t};\theta_{\pi_i})
\nabla_{\theta_{Q_\pi}^g}Q_{\theta_{Q_\pi}^g}({s}_t,{a}_t)\Big]}_{\text{\footnotesize MADDPG}}\\
&+\underbrace{\mathbb{E}_{s_{i,t},a_{i,t}\sim \mathcal{B}}\Big[\nabla_{\theta_{\pi_i}}\pi_i({a}_{i,t}|{s}_{i,t};\theta_{\pi_i})
\nabla_{\theta_{Q_{\pi_i}}}Q_{\theta_{Q_{\pi_i}}}({s}_{i,t},{a}_{i,t})\Big]}_{\text{\footnotesize DDPG}},
\end{aligned}
\label{eq21}
\end{equation} 
where $\mathcal{B}$ is the experience extraction replay buffer, $Q_{\theta_{Q_\pi}^g}({s}_t,{a}_t)$ is the global action value and $Q_{\theta_{Q_{\pi_i}}}({s}_{i,t},{a}_{i,t})$ is the local action value of agent $i$, respectively.

Besides, the global action value $Q_{\theta_{Q_\pi}^g}({s}_t,{a}_t)$ and the local action value $Q_{\theta_{Q_{\pi_i}}}({s}_{i,t},{a}_{i,t})$ are calculated by the global \itshape Critic \upshape network and local \itshape Critic \upshape network, respectively. Correspondingly, the mean-squared Bellman error function of the global critic network $L(\theta_{Q_\pi}^g)$ can be defined as
\begin{equation}
\setcounter{equation}{22}
\begin{aligned}
L(\theta_{Q_\pi}^g) = \mathbb{E}_{s_t,a_t,r_t,s_{t+1}}\Big[\Big(Q_{\theta_{Q_\pi}^g}({s}_t,{a}_t)-y_{t}^g\Big)^2\Big],
\label{eq17}
\end{aligned}
\end{equation}
where $y_t^g=r_t+\gamma\Big(Q_{\theta_{Q_{\pi'}}^g}({s}_t',{a}_t')\Big)$ is the global target, and $Q_{\theta_{Q_{\pi'}}^g}({s}_t',{a}_t')$ is the global value.

And the mean-squared Bellman error function of the local \itshape Critic \upshape network can be defined as
\begin{equation}
\setcounter{equation}{23}
\begin{aligned}
L(\theta_{Q_{\pi_i}}) = \mathbb{E}_{s_{t},a_{t},r_{t},s_{t+1}}\Big[\Big(Q_{\theta_{Q_{\pi_i}}}({s}_{i,t},{a}_{i,t})-y_{i,t}^l\Big)^2\Big],
\label{eq18}
\end{aligned}
\end{equation}
where $y_{i,t}^l=r_{i,t}+\gamma\Big(Q_{\theta_{Q_{{\pi_i}'}}}({s}_{i,t}',{a}_{i,t}')\Big)$ is the local target, and $Q_{\theta_{Q_{{\pi_i}'}}}({s}_{i,t}',{a}_{i,t}')$ is the local value.

Finally, to ensure that the target network remains stable throughout the iterative process, a soft update is carried out with the update rate $\tau\ll 1$. The local \itshape target Actor \upshape network and local \itshape target Critic \upshape network are
\begin{equation}
\setcounter{equation}{24}
\begin{split}
\left \{
\begin{array}{ll}
    {\theta_{\pi_i'}}\leftarrow\tau{\theta_{\pi_i'}}+(1-\tau)\theta_{\pi_i},\\
    {\theta_{Q_{\pi_i'}}} \leftarrow \tau{\theta_{Q_{\pi_i'}}} + (1-\tau)\theta_{Q_{\pi_i}}.
\end{array}
\right.
\label{eq19}
\end{split}
\end{equation}
\begin{figure*}[t]
    \centering
    \includegraphics[width=5.5in]{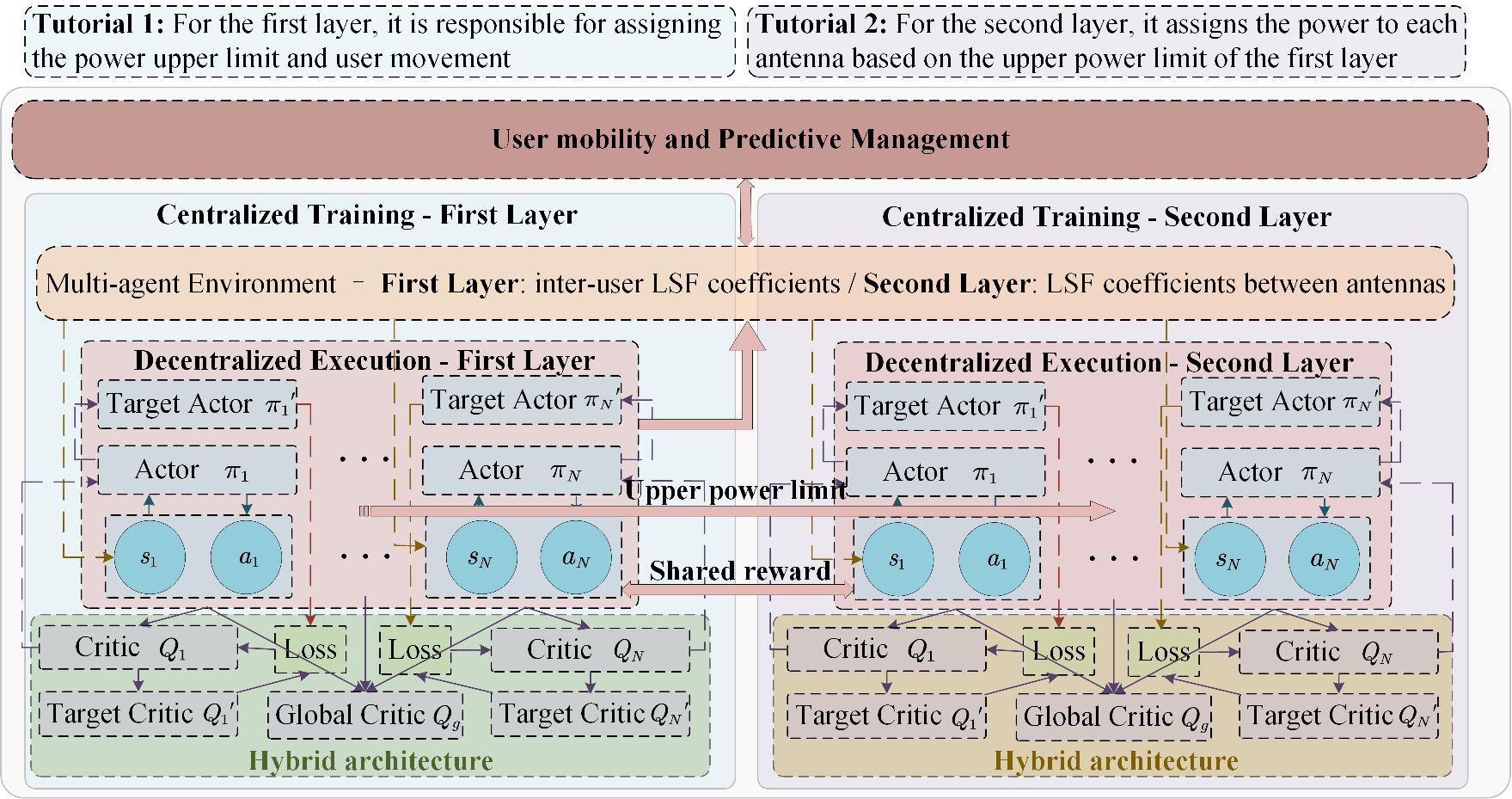}
    \caption{Illustration of the double-layer power control architecture.}
    \label{fig3}
\end{figure*}

And the global \itshape target Critic \upshape network is
\begin{equation}
\setcounter{equation}{25}
\begin{split}
{\theta_{Q_{\pi'}}^g} \leftarrow \tau{\theta_{Q_{\pi'}}^g} + (1-\tau)\theta_{Q_\pi}^g.
\label{eq20}
\end{split}
\end{equation}

The corresponding procedure of the MIMO-MADDPG is summarized in $\textbf{Algorithm 1}$. Moreover, by analyzing different signal architectures in Fig. 2, we can observe that the training amount of the network is closely related to the following parameters: $M$, $K$, $N_r$, $N_s$, and hyperparameters. Firstly, we analyze the computational complexity of MADDPG, which consists of two parts: MARL-based neural network and reward function. The complexity of MARL-based neural network is $\mathcal{O}(\sum_{a=1}^{A}MK^{3}Q_a^{2}d_a+\sum_{c=1}^{C}(MK^{2}+Kd_a)Q_c^{2})$, where $A$ and $C$ are the number of hidden layers at the \itshape Actor \upshape network and \itshape Critic \upshape network, respectively, $Q_a$ denotes the output size of the $a$-th layer or the input size of the next layer, and $d_a$ denotes the dimension of the output action. And the complexity of reward function is $\mathcal{O}(M^{2}K^{2}+MKN_rN_s)$.
In addition, we analyze the computational complexity of PES-MADDPG and DE-MADDPG.
Compared with the computational complexity of MADDPG, the priority experience extraction network and global Critic network are added respectively, and the corresponding complexity of both is $\mathcal{O}(K^{2}N_{batch}^{\mathcal{B}})$ and $\mathcal{O}(\sum_{c=1}^{C}(MK^{2}+Kd_a)Q_c^{2})$, respectively. Finally, based on the above analysis, we can obtain the computational complexity of the proposed MIMO-MADDPG and FL-MADDPG, both of which are composed of three parts: MARL-based neural network, reward function, and additional system network, as shown in Table \uppercase\expandafter{\romannumeral1}.
\begin{algorithm}[t]
  \caption{MIMO-MADDPG}
  \label{alg::conjugateGradient}
  \begin{algorithmic}[1]
      \State
        \textbf{Initialize} The global and local experience extraction pool $\mathcal{B}_g$ and $\mathcal{B}_l$, the number of global and local experiences extracted from $\mathcal{B}_g$ and $\mathcal{B}_l$ at a time: $\mathcal{M}_g$ and $\mathcal{M}_l$.
      \For {episode = 1 to max-episodes}
        \For {step = 1 to max-steps}
            \State Get initial state ${s}_{t}$
            \State Actor network determines the assigned action ${a}_{i,t}$
            \State Obtain the actual expected rewards $r_{t}$
            \State Get the next state $s_{t+1}$ after the agent interacts with the environment
            \State Calculate the target evaluation network loss $loss_{t}$
            \State Initialize the priority $Pr_{t} = 0$ and the experience extraction training times $n_{t} = 0$
            \State Store $<{s}_{t}, {a}_{t}, {r}_{t}, {s}_{t+1}, loss_t, n_t, Pr_t>$ to the experience pool $\mathcal{D}_g$ and $\mathcal{D}_l$
            \State Update the priority $Pr_t$, and fill $\mathcal{B}_g$ and $\mathcal{B}_l$
            \If {update the global network}
                \State Sample $\mathcal{M}_g$ experiences from $\mathcal{B}_g$ randomly
                \State Calculate the global critic value $L(Q_{\theta_{Q_\pi}^g})$
                \State Update the weights of global critic network
                \State Update the global target critic network ${\theta_{Q_{\pi'}}^g}$: ${\theta_{Q_{\pi'}}^g} \leftarrow \tau{\theta_{Q_{\pi'}}^g} + (1-\tau)\theta_{Q_\pi}^g$
            \EndIf
            \If {update the local network}
              \For {agent $i$ = 1 to $K$}
                \State Sample $\mathcal{M}_l$ experiences from $\mathcal{B}_l$ randomly
                \State Calculate the local critic value
                \State Update the weights of local critic network
                \State Calculate the policy gradient
                \State Update the local target critic network ${\theta_{Q_{\pi_i'}}}$: ${\theta_{Q_{\pi_i'}}} \leftarrow \tau{\theta_{Q_{\pi_i'}}} + (1-\tau)\theta_{Q_{\pi_i}}$
                \State Update the local target actor network ${\theta_{{\pi_i'}}}$: ${\theta_{\pi_i'}}\leftarrow\tau{\theta_{\pi_i'}}+(1-\tau)\theta_{\pi_i}$
              \EndFor
            \EndIf
        \EndFor
      \EndFor
  \end{algorithmic}
\end{algorithm}
\subsection{Double-Layer Power Control}
In this subsection, we introduce a double-layer power control architecture based on the LSF coefficient $\mathbf{B}_{mk} \in \mathbb{C}^{N_r\times N_s}$ between antennas in CF XL-MIMO systems. This architecture differs from the conventional single-layer power control architecture that only considers the inter-user LSF coefficient $\beta_{mk}$ and treats all antennas under the same agent as a whole.
In the design of double-layer architecture, we define UEs and antennas as heterogeneous agents deployed under different architectures, such as dynamic architecture and static architecture. The corresponding double-layer power control architecture is shown in Fig. 3.

\itshape 1) First Layer: Dynamic Power Control Architecture \upshape

In the first layer, all $K$ UEs are regarded as agents deployed in dynamic scenarios, and all antennas belonging to the same UE are analyzed as a whole. Based on this, the state and action of the first layer can be defined as $s_{t,(1)}=[\beta_{1},\ldots,\beta_{K}]$ in conjunction with $\beta_{k}=\sum_{m=1}^{M}\beta_{mk}$ and $a_{t,(1)}=[a_{t,(1)}^p;a_{t,(1)}^d;a_{t,(1)}^\theta]$, respectively.
And the rewards of the first layer $r_{t,(1)}$ can be calculated using equation (14). The primary design purpose of the layer is to provide the upper limit of power for the second layer and to enable the dynamic movement of all agents.

Then, the policy gradient can be modeled based on the single-layer architecture as 
\begin{equation}
\setcounter{equation}{26}
\begin{split}
\begin{aligned}
\nabla_{\theta_{\pi_i,(1)}}J(\theta_{\pi_i,(1)})=
\underbrace{\mathbb{E}_{s_{t,(1)},a_{t,(1)}\sim \mathcal{B}_{(1)}}\bigg[\nabla_{\theta_{\pi_i,(1)}}\pi_i\Big({a}_{i,t,(1)}\Big|{s}_{i,t,(1)};{\theta_{\pi_i,(1)}}\Big)
\nabla_{{\theta_{Q_\pi,(1)}^g}}Q_{\theta_{Q_\pi,(1)}^g}\Big({s}_{t,(1)},{a}_{t,(1)}\Big)\bigg]}_{\text{\footnotesize MADDPG}}\\
+\underbrace{\mathbb{E}_{s_{i,t,(1)},a_{i,t,(1)}\sim \mathcal{B}_{(1)}}\bigg[\nabla_{\theta_{\pi_i,(1)}}\pi_i\Big({a}_{i,t,(1)}\Big|{s}_{i,t,(1)};{\theta_{\pi_i,(1)}}\Big)
\nabla_{\theta_{Q_{\pi_i},(1)}}Q_{\theta_{Q_{\pi_i},(1)}}\Big({s}_{i,t,(1)},{a}_{i,t,(1)}\Big)\bigg]}_{\text{\footnotesize DDPG}}.
\end{aligned}
\end{split}
\end{equation}

Correspondingly, the mean-squared Bellman error function of the global \itshape Critic \upshape network in the first layer $L_{(1)}^g$ can be defined as
\begin{equation}
\setcounter{equation}{27}
\begin{split}
L_{(1)}^g = \mathbb{E}_{\mathcal{B}_{(1)}}\Big[\Big(Q_{\theta_{Q_{\pi},(1)}^g}({s}_{t,(1)},{a}_{t,(1)})-y_{t,(1)}^g\Big)^2\Big],
\label{eq22}
\end{split}
\end{equation}
where $y_{t,(1)}^g=r_{t,(1)}+\gamma(Q_{\theta_{Q_{\pi'},(1)}^g}({s}_{t,(1)}',{a}_{t,(1)}'))$ is the global target, and $Q_{\theta_{Q_{\pi'},(1)}^g}({s}_{t,(1)}',{a}_{t,(1)}')$ is the global value.

Furthermore, the mean-squared Bellman error function of the local \itshape Critic \upshape network of agent $i$ in the first layer $L_{i,(1)}^l$ can be defined as
\begin{equation}
\setcounter{equation}{28}
\begin{split}
L_{i,(1)}^l = \mathbb{E}\Big[\Big(Q_{\theta_{Q_{\pi_i},(1)}}({s}_{i,t,(1)},{a}_{i,t,(1)})-y_{i,t,(1)}^l\Big)^2\Big],
\label{eq23}
\end{split}
\end{equation}
where $y_{i,t,(1)}^l=r_{i,t,(1)}+\gamma(Q_{\theta_{Q_{\pi_i'},(1)}}({s}_{i,t,(1)}',{a}_{i,t,(1)}'))$ is the local target, and $Q_{\theta_{Q_{\pi_i'},(1)}}({s}_{i,t,(1)}',{a}_{i,t,(1)}')$ is the local value.

Finally, in order to ensure that the local \itshape target \upshape network remains
stable in the iterative process, the soft update is carried out with $\tau\ll 1$. The local \itshape target \upshape  network of the first layer is
\begin{equation}
\setcounter{equation}{29}
\begin{split}
\left \{
\begin{array}{ll}
    {\theta_{\pi_i',(1)}}\leftarrow\tau{\theta_{\pi_i',(1)}}+(1-\tau)\theta_{\pi_i,(1)},\\
    {\theta_{Q_{\pi_i'},(1)}} \leftarrow \tau{\theta_{Q_{\pi_i'},(1)}} + (1-\tau)\theta_{Q_{\pi_i},(1)}.
\end{array}
\right.
\label{eq24}
\end{split}
\end{equation}

And the global \itshape target \upshape network of the first layer is
\begin{equation}
\setcounter{equation}{30}
\begin{split}
{\theta_{Q_{\pi'},(1)}^g} \leftarrow \tau{\theta_{Q_{\pi'},(1)}^g} + (1-\tau)\theta_{Q_{\pi},(1)}^g.
\label{eq25}
\end{split}
\end{equation}
\begin{table*}[t]
  \centering
  \fontsize{9}{11.4}\selectfont
  \caption{Comparison of relevant algorithms with the proposed algorithm.}
  \label{Paper_comparison}
    \begin{tabular}{ !{\vrule width1.2pt}  m{2.6 cm}<{\centering} !{\vrule width1.2pt}   m{1.45 cm}<{\centering} !{\vrule width1.2pt}  m{2 cm}<{\centering} !{\vrule width1.2pt}  m{1.75 cm}<{\centering}  !{\vrule width1.2pt} m{1.85 cm}<{\centering} !{\vrule width1.2pt} m{2.05 cm}<{\centering} !{\vrule width1.2pt} m{1.65 cm}<{\centering} !{\vrule width1.2pt}}

    \Xhline{1.2pt}
        \rowcolor{gray!30} \bf Methods  &  \bf Near-field & \bf Generalization ability &  \bf Convergence stability &  \bf Instantaneity &  \bf Predictive Management & \bf Low-delay interaction \cr
    \Xhline{1.2pt}

        \bf MADDPG   & \makecell[c]{\XSolidBrush} & \makecell[c]{\XSolidBrush} & \makecell[c]{\Checkmark} & \makecell[c]{\XSolidBrush} & \makecell[c]{\XSolidBrush} & \makecell[c]{\XSolidBrush} \cr\hline
        \bf FL-MADDPG  & \makecell[c]{\XSolidBrush} & \makecell[c]{\Checkmark} & \makecell[c]{\XSolidBrush} & \makecell[c]{\XSolidBrush} & \makecell[c]{\XSolidBrush} & \makecell[c]{\Checkmark} \cr\hline
        \bf DE-MADDPG & \makecell[c]{\XSolidBrush} & \makecell[c]{\XSolidBrush} & \makecell[c]{\Checkmark} & \makecell[c]{\XSolidBrush} & \makecell[c]{\XSolidBrush} & \makecell[c]{\Checkmark} \cr\hline
        \bf PES-MADDPG  & \makecell[c]{\XSolidBrush} & \makecell[c]{\XSolidBrush} & \makecell[c]{\Checkmark}  & \makecell[c]{\XSolidBrush} & \makecell[c]{\XSolidBrush} & \makecell[c]{\Checkmark} \cr\hline
        \bf MIMO-MADDPG  & \makecell[c]{\Checkmark} & \makecell[c]{\Checkmark} & \makecell[c]{\Checkmark} & \makecell[c]{\Checkmark} & \makecell[c]{\Checkmark} & \makecell[c]{\Checkmark} \cr\hline
    \Xhline{1.2pt}
    \end{tabular}
  \vspace{0cm}
\end{table*}
\itshape 2) Second Layer: Static Power Control Architecture \upshape

In contrast to the first layer architecture, the second layer considers all antennas under each UE as independent agents deployed in static scenarios, eliminating the need for assigned agents to perform mobile actions. Additionally, based on the constraints of the upper power limit, the LSF coefficients between antennas are used to complete the power allocation of antennas. Therefore, the state of the second layer can be defined as $s_{t,(2)}=[\mathbf{B}_{1,1},\mathbf{B}_{1,2},\ldots,\mathbf{B}_{K,N_s};a_{kn_s}^p]$ with $\mathbf{B}_{k,n_s}=\sum_{m=1}^{M}\sum_{r=1}^{N_r}\mathbf{B}_{mk}^{r,n_s},k=[1,\ldots,K],n_s=[1,\ldots,N_s]$, the upper power limit
\begin{equation}
\setcounter{equation}{31}
\begin{split}
a_{kn_s}^p=[\underbrace{a_{1,t,(1)}^p,\ldots,a_{1,t,(1)}^p}_{N_s},\ldots,\underbrace{a_{K,t,(1)}^p,\ldots,a_{K,t,(1)}^p}_{N_s}],
\label{eq26}
\end{split}
\end{equation}
and the action of the second layer can be defined as $a_{t,(2)}=a_{t,(2)}^p$, respectively. It is worth noting that the reward value is shared in the double-layer architecture. Therefore, we can calculate the reward of the second layer $r_{t,(2)} \in \mathbb{R}^{KN_s\times 1}$ using the reward of the first layer $r_{t,(1)} \in \mathbb{R}^{K\times 1}$ as
\begin{equation}
\setcounter{equation}{32}
\begin{split}
r_{t,(2)}=\Big[\underbrace{\frac{r_{1,t,(1)}}{N_s},\ldots,\frac{r_{1,t,(1)}}{N_s}}_{N_s},\ldots,\underbrace{\frac{r_{K,t,(1)}}{N_s},\ldots,\frac{r_{K,t,(1)}}{N_s}}_{N_s}\Big].
\label{eq27}
\end{split}
\end{equation}

Additionally, the policy gradient of the second layer $\nabla_{\theta_{\pi_i,(2)}}J(\theta_{\pi_i,(2)})$ can be modeled as 
\begin{equation}
\setcounter{equation}{33}
\begin{split}
\begin{aligned}
\nabla_{\theta_{\pi_i,(2)}}J(\theta_{\pi_i,(2)})=
\underbrace{\mathbb{E}_{s_{t,(2)},a_{t,(2)}\sim \mathcal{B}_{(2)}}\bigg[\nabla_{\theta_{\pi_i,(2)}}\pi_i\Big({a}_{i,t,(2)}\Big|{s}_{i,t,(2)};{\theta_{\pi_i,(2)}}\Big)
\nabla_{\theta_{Q_\pi,(2)}^g}Q_{\theta_{Q_\pi,(2)}^g}\Big({s}_{t,(2)},{a}_{t,(2)}\Big)\bigg]}_{\text{\footnotesize MADDPG}}\\
+\underbrace{\mathbb{E}_{s_{i,t,(2)},a_{i,t,(2)}\sim \mathcal{B}_{(2)}}\bigg[\nabla_{\theta_{\pi_i,(2)}}\pi_i\Big({a}_{i,t,(2)}\Big|{s}_{i,t,(2)};{\theta_{\pi_i,(2)}}\Big)
\nabla_{\theta_{Q_{\pi_i},(2)}}Q_{\theta_{Q_{\pi_i},(2)}}\Big({s}_{i,t,(2)},{a}_{i,t,(2)}\Big)\bigg]}_{\text{\footnotesize DDPG}}.
\end{aligned}
\end{split}
\end{equation}

Correspondingly, the mean-squared Bellman error function of the global \itshape Critic \upshape network in the second layer $L_{(2)}^g$ can be defined as
\begin{equation}
\setcounter{equation}{34}
\begin{split}
L_{(2)}^g = \mathbb{E}_{\mathcal{B}_{(2)}}\Big[\Big(Q_{\theta_{Q_{\pi},(2)}^g}({s}_{t,(2)},{a}_{t,(2)})-y_{t,(2)}^g\Big)^2\Big],
\label{eq29}
\end{split}
\end{equation}
where $y_{t,(2)}^g=r_{t,(2)}+\gamma(Q_{\theta_{Q_{\pi'},(2)}^g}({s}_{t,(2)}',{a}_{t,(2)}'))$ and $Q_{\theta_{Q_{\pi'},(2)}^g}({s}_{t,(2)}',{a}_{t,(2)}')$ are the global target and value.

Furthermore, the mean-squared Bellman error function of the local \itshape Critic \upshape network of agent $i$ in the second layer $L_{i,(2)}^l$ can be defined as
\begin{equation}
\setcounter{equation}{35}
\begin{split}
L_{i,(2)}^l = \mathbb{E}\Big[\Big(Q_{\theta_{Q_{\pi_i},(2)}}({s}_{i,t,(2)},{a}_{i,t,(2)})-y_{i,t,(2)}^l\Big)^2\Big],
\label{eq30}
\end{split}
\end{equation}
where $y_{i,t,(2)}^l=r_{i,t,(2)}+\gamma(Q_{\theta_{Q_{\pi_i'},(2)}^g}({s}_{i,t,(2)}',{a}_{i,t,(2)}'))$ is the local target, and $Q_{\theta_{Q_{\pi_i'},(2)}^g}({s}_{i,t,(2)}',{a}_{i,t,(2)}')$ is the local value.
\begin{table}[t]
  \centering
\fontsize{9}{11}\selectfont
  \caption{The model structure in our experiments.}
  \label{paper}
\begin{tabular}{ccc}
\toprule
\bf Parameters &  \bf Size \\
\midrule
1st hidden layer & 128, Leaky Relu (0.01)\\
2nd hidden layer & 64, Leaky Relu (0.01) \\
Discounted factor $\gamma$ & 0.99 \\
Experience pool size $N_{batch}^{\mathcal{D}}$ & 1024 \\
Experience extraction pool size $N_{batch}^{\mathcal{B}}$ & 512 \\
Maximum gradient clipping value $\xi$ & 0.5 \\
Soft update rate $\tau$ & 0.01 \\
Learning rate & 0.01 \\
\bottomrule
\end{tabular}
\end{table}

Finally, similar to the first layer architecture, to ensure stable network convergence, the soft update is carried out with $\tau\ll 1$. The local \itshape target \upshape  network of the second layer is
\begin{equation}
\setcounter{equation}{36}
\begin{split}
\left \{
\begin{array}{ll}
    {\theta_{\pi_i',(2)}}\leftarrow\tau{\theta_{\pi_i',(2)}}+(1-\tau)\theta_{\pi_i,(2)},\\
    {\theta_{Q_{\pi_i'},(2)}} \leftarrow \tau{\theta_{Q_{\pi_i'},(2)}} + (1-\tau)\theta_{Q_{\pi_i},(2)}.
\end{array}
\right.
\label{eq31}
\end{split}
\end{equation}

And the global \itshape target \upshape network of the second layer is
\begin{equation}
\setcounter{equation}{37}
\begin{split}
{\theta_{Q_{\pi',(2)}}^g} \leftarrow \tau{\theta_{Q_{\pi',(2)}}^g} + (1-\tau)\theta_{Q_{\pi,(2)}}^g.
\label{eq32}
\end{split}
\end{equation}

\subsection{Comparative Analysis of Architectures}
In this subsection, we analyze the ability of various MARL-based methods using different architectures to adapt to varying system conditions and present the corresponding comparison results in Table \uppercase\expandafter{\romannumeral2}. It is noteworthy that the FL-MADDPG algorithm proposed in \cite{[2]} employs fuzzy agents to reduce computational complexity, thereby improving the realizability of CF XL-MIMO systems. As for the remaining algorithms, i.e., the conventional MARL-based algorithms proposed in \cite{[6],[7],[9]}, as well as the proposed MIMO-MADDPG algorithm, they all improve the realizability by optimizing the convergence rate, which is more conducive to consolidating the convergence stability of the designed algorithms.

Moreover, all of the conventional MARL-based algorithms only take advantage of user mobility in high-speed mobile scenarios. In contrast, the proposed MIMO-MADDPG algorithm has an additional predictive management architecture that not only limits the movement of all agents and avoids agents getting stuck in local optimal solutions for a long time, but also adjusts the transmit power accordingly based on the predicted channel conditions. This also shows that the proposed method has higher advantages in adapting to instantaneous system scenarios and high mobile scenarios with changing channel conditions.
\begin{figure}[t]
	\centering
	\begin{minipage}[t]{0.48\textwidth}
	\centering
    \includegraphics[scale=0.5]{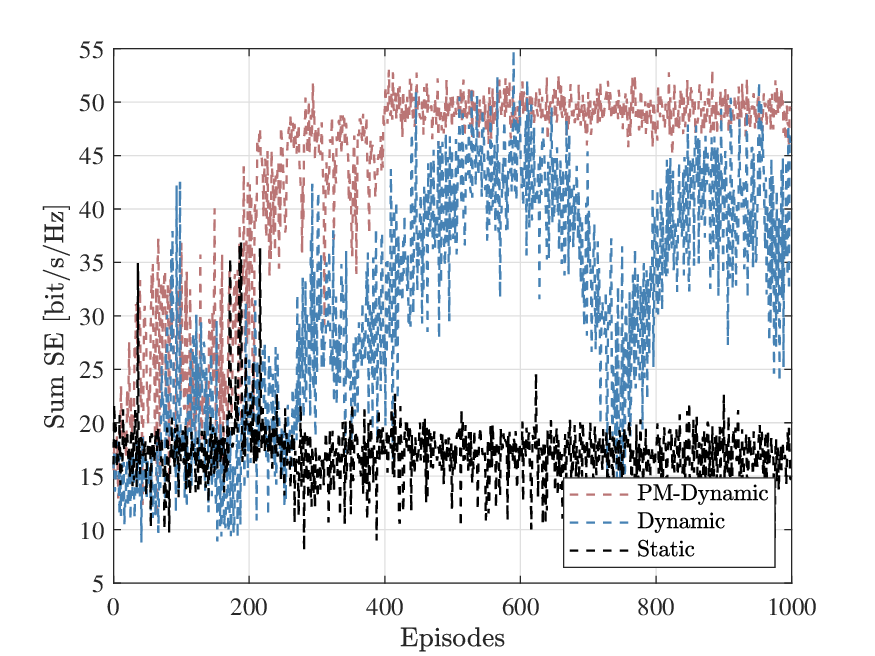}
    \caption{Sum SE over different scenarios under MADDPG for MR combining with $M=9$, $K=6$, $N_r=N_{H_r}\times N_{V_r}=81$, $N_s=N_{H_s}\times N_{V_s}=9$, and $\Delta_s = \Delta_r = \lambda/3$.}
	\end{minipage}
    \hfill%
	\begin{minipage}[t]{0.48\textwidth}
	\centering
    \includegraphics[scale=0.5]{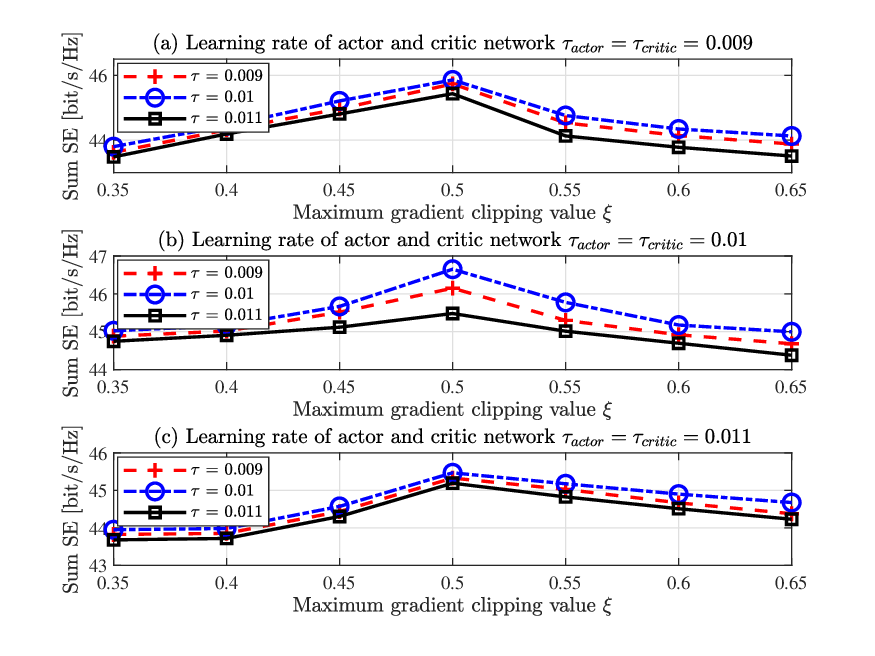}
    \caption{The effect of different combinations of hyperparameters on system performance for MR combining with $M=9$, $K=6$, $N_r=N_{H_r}\times N_{V_r}=81$, $N_s=N_{H_s}\times N_{V_s}=9$, and $\Delta_r = \lambda/3$.}
	\end{minipage}
\end{figure}
\begin{figure}[t]
    \hfill%
	\centering
	\begin{minipage}[t]{0.48\textwidth}
	\centering
    \includegraphics[scale=0.5]{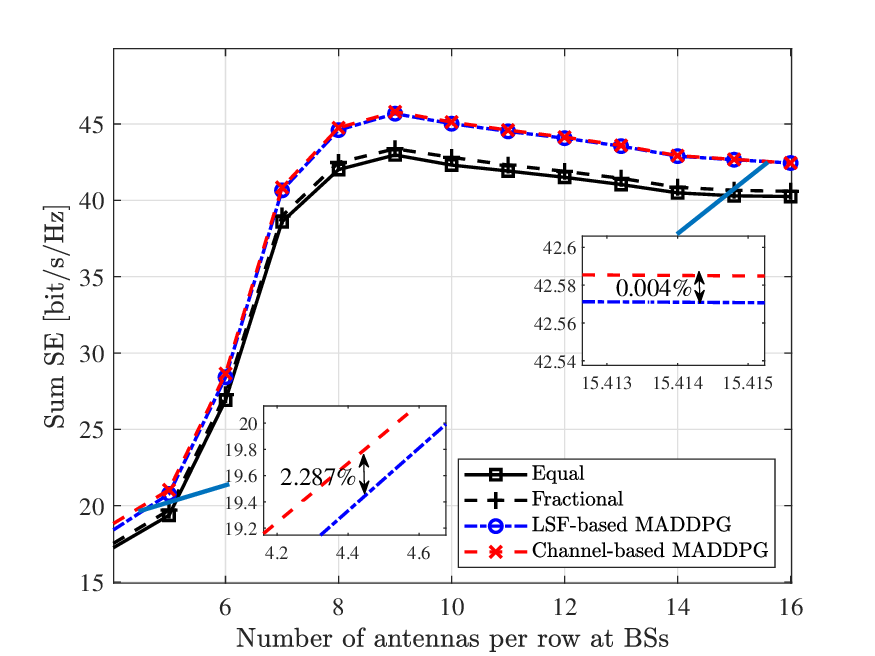}
    \caption{The achievable sum SE over different power control schemes against the number of antennnas at each BS with $M=9$, $K=6$, $N_s=N_{H_s}\times N_{V_s}=9$, and $\Delta_s = \Delta_r = \lambda/3$.}
	\end{minipage}
    \hfill%
	\centering
	\begin{minipage}[t]{0.48\textwidth}
	\centering
    \includegraphics[scale=0.5]{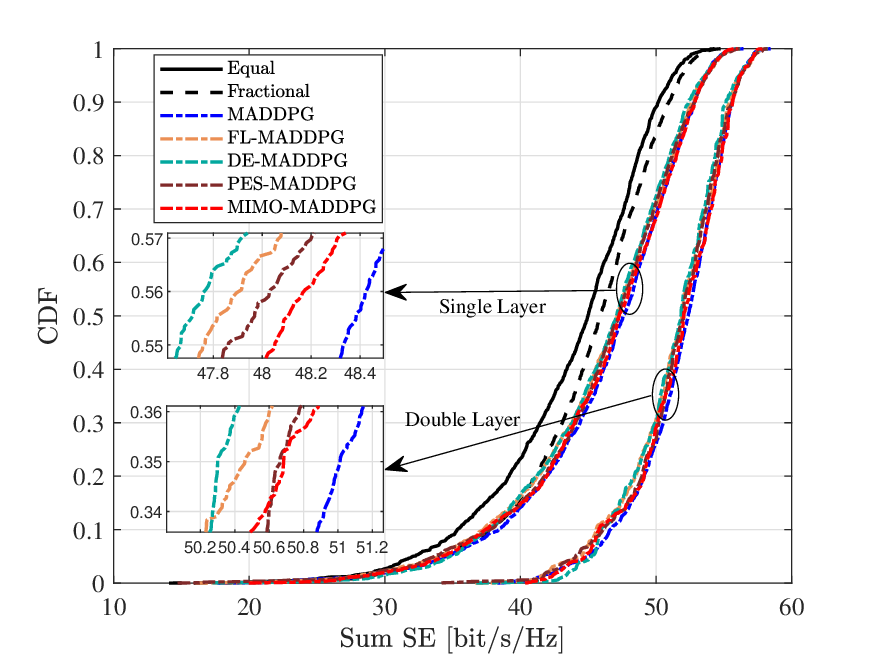}
    \caption{CDF of Sum SE over different algorithm architectures for MR combining with $M=9$, $K=6$, $N_r=N_{H_r}\times N_{V_r}=81$, $N_s=N_{H_s}\times N_{V_s}=9$, and $\Delta_s = \Delta_r = \lambda/3$.}
	\end{minipage}
    \hfill%
	\begin{minipage}[t]{0.48\textwidth}
	\centering
    \includegraphics[scale=0.5]{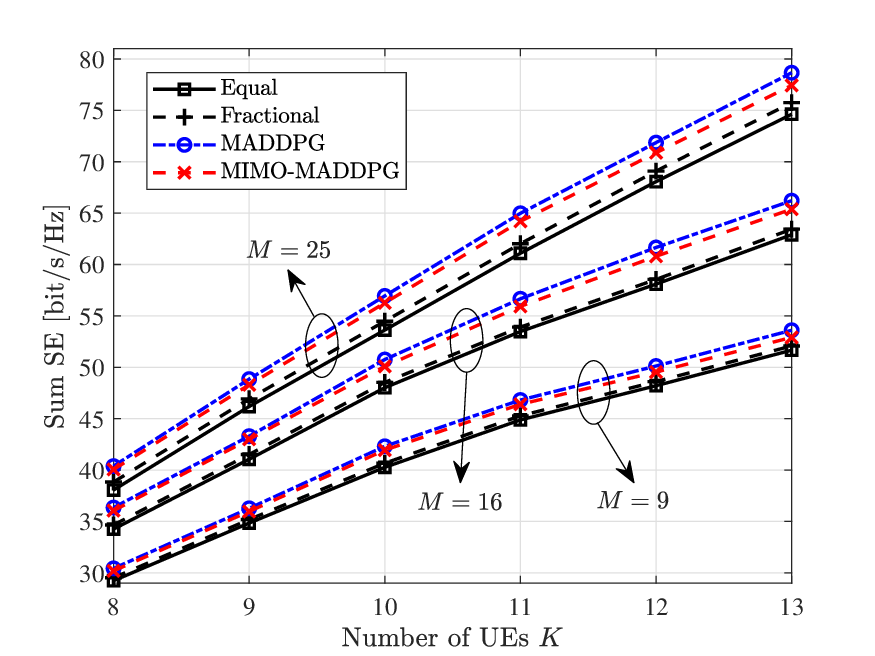}
    \caption{The achievable sum SE over different MARL-based algorithms against the number of UEs and BSs with $N_r=N_{H_r}\times N_{V_r}=81$, $N_s=N_{H_s}\times N_{V_s}=9$, and $\Delta_s = \Delta_r = \lambda/3$.}
	\end{minipage}
\end{figure}
\begin{figure}[t]
    \hfill%
	\begin{minipage}[t]{0.48\textwidth}
	\centering
    \includegraphics[scale=0.5]{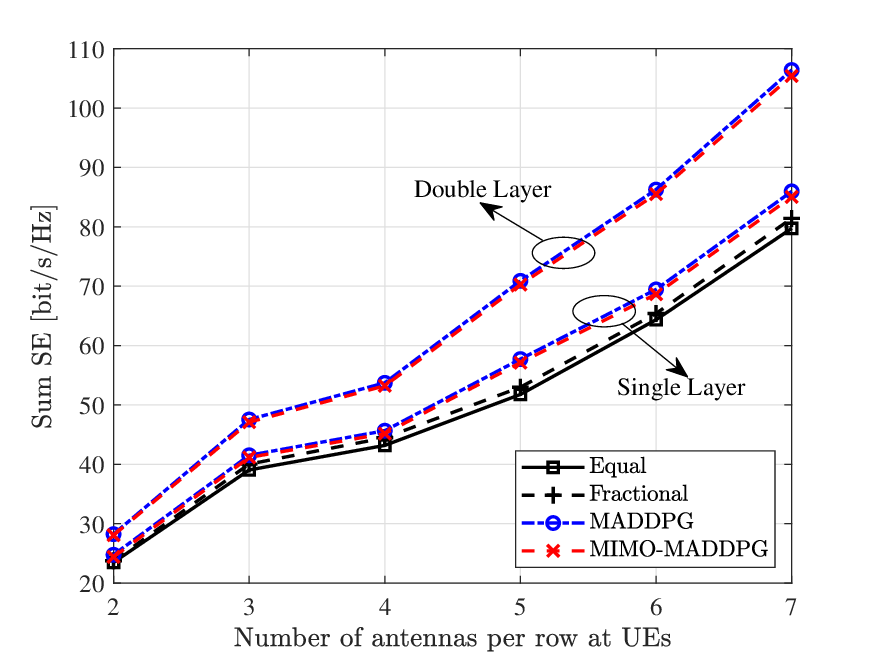}
    \caption{Sum SE against the number of antennas per row at UEs for MR combining with $M=9$, $K=6$, $N_r=N_{H_r}\times N_{V_r}=100$, and $\Delta_s = \Delta_r = \lambda/3$.}
	\end{minipage}
	\begin{minipage}[t]{0.48\textwidth}
	\centering
    \includegraphics[scale=0.5]{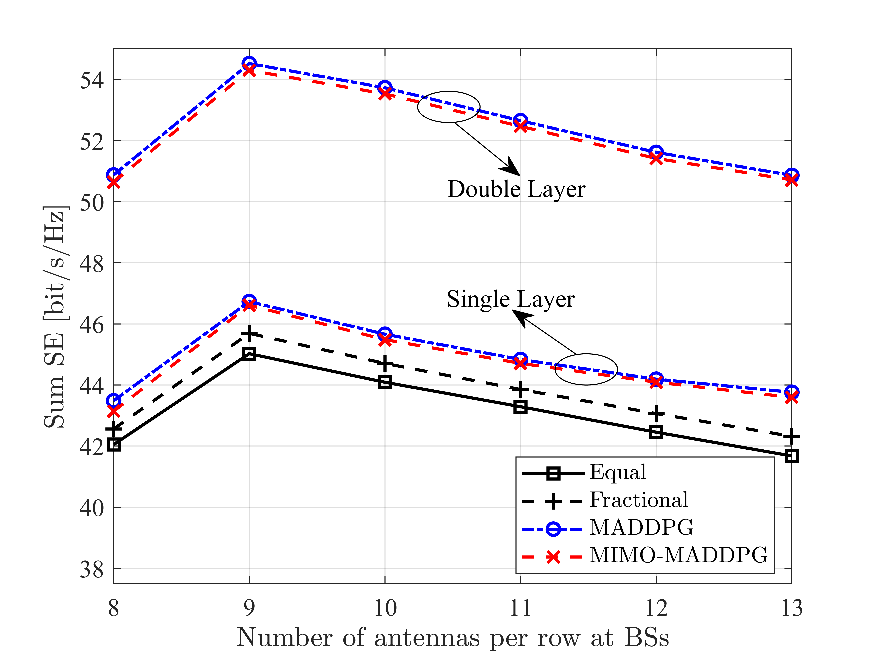}
    \caption{Sum SE against the number of antennas per row at BSs for MR combining with $M=9$, $K=6$, $N_s=N_{H_s}\times N_{V_s}=16$, and $\Delta_s = \Delta_r = \lambda/3$.}
	\end{minipage}
\end{figure}
\begin{figure}[t]
	\centering
	\begin{minipage}[t]{0.48\textwidth}
	\centering
    \includegraphics[scale=0.5]{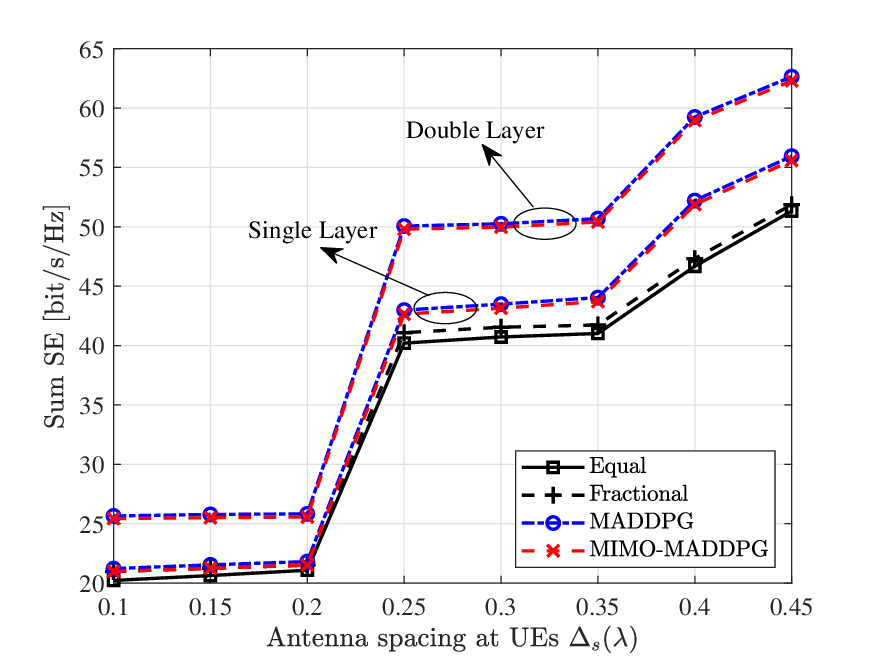}
    \caption{Sum SE against antenna spacing at UEs for MR combining with $M=9$, $K=6$, $N_r=N_{H_r}\times N_{V_r}=100$, $N_s=N_{H_s}\times N_{V_s}=16$, and $\Delta_r = \lambda/4$.}
	\end{minipage}
    \hfill%
	\begin{minipage}[t]{0.48\textwidth}
	\centering
    \includegraphics[scale=0.5]{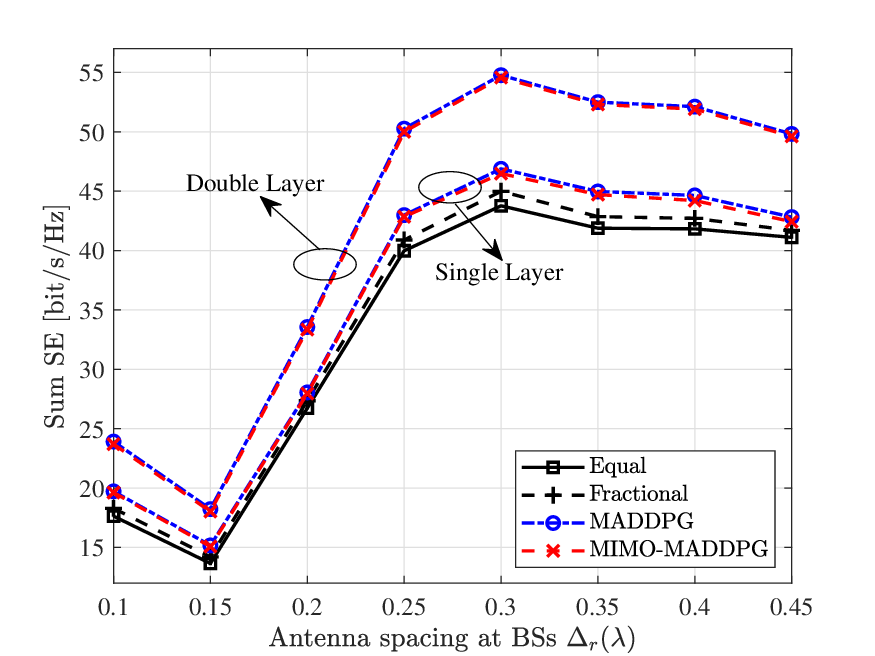}
    \caption{Sum SE against antenna spacing at BSs for MR combining with $M=9$, $K=6$, $N_r=N_{H_r}\times N_{V_r}=100$, $N_s=N_{H_s}\times N_{V_s}=16$, and $\Delta_s = \lambda/4$.}
	\end{minipage}
\end{figure}
\section{Numerical Results}
In this paper, we investigate a CF XL-MIMO system where all BSs and UEs are uniformly distributed in a $1\times 1$ $\text{km}^2$ area, as depicted in Fig. 1, utilizing a wrap-around scheme, which is more in line with the realistic scene. Based on this scheme, we have carefully placed the BSs to avoid any overlapping coverage areas that could lead to self-interference or cross-interference. Moreover, we ensure in the simulation setup that there is a distance of at least 200 meters (over 50 meters) between adjacent BSs so that the correlation with the shadow terms associated with two different BSs is negligible.
Then, we consider a 20 MHz communication bandwidth and set the noise power to $\sigma^2=-69$ dBm. All UEs transmit with a transmission power of no more than 200 mW. In the proposed algorithms, both the \itshape Actor \upshape and \itshape Critic \upshape network consist of two hidden layers, and the model structure and experimental details are shown in Table \uppercase\expandafter{\romannumeral3}.
Furthermore, we set up the experimental environment and complete the simulation with PyTorch, and the training works are executed with an Nvidia GeForce GTX 3060 Graphics Processing Unit.

Moreover, considering that the wireless environment studied is dynamically changing, we need to monitor the MARL model in real-time during the training process and compare the performance of the Fractional algorithm in simulation as the baseline algorithm to determine whether the trained MARL model is outdated. Since the emergence of outdated models can have a serious impact on system performance, we introduce an online learning mechanism into the network architecture, so that agents can constantly learn and update their policies based on the current wireless conditions. This helps ensure that the trained MARL model remains up-to-date in an ever-changing wireless environment and achieves superior system performance.
\subsection{Effects of User Mobility and Predictive Management}
We first investigate the effects of user mobility and predictive management. Fig. 4 shows the achievable sum SE for three different scenarios based on the MADDPG algorithm investigated in this paper over MR combining. We notice that PM-Dynamic scenarios, as well as common Dynamic scenarios, undoubtedly achieve higher SE compared to Static scenarios, with SE performance improving by 60.32\% and 45.23\%, respectively. The reason behind that is Dynamic scenarios fully adopt the characteristics of the state transition mechanism in MARL-based algorithms, in which all agents always move towards a better target point.
More important, the proposed user mobility and predictive management method is efficient to improve the achievable sum SE performance by 27.45\% for PM-Dynamic scenarios compared to common Dynamic scenarios.
\subsection{Sensitivity of Key Hyperparameters}
We investigate the sensitivity of key hyperparameters, e.g., the learning rate of \itshape Actor \upshape network $\tau_{actor}$, learning rate of \itshape Critic \upshape network $\tau_{critic}$, soft update rate $\tau$, and maximum gradient clipping value $\xi$. Fig. 5 illustrates the effect of different combinations of hyperparameters on system performance. By comparing different hyperparameter combinations, we can notice that the hyperparameters during MARL training are sensitive to the environment, and different combinations of hyperparameters have a great impact on the system performance. Moreover, it can be found that in the specific mMIMO scenario studied in this paper, in order to obtain better spectral efficiency performance, the combination of key hyperparameters can be set to the learning rate of \itshape Actor \upshape network $\tau_{actor}=0.01$, the learning rate of \itshape Critic \upshape network $\tau_{critic}=0.01$, soft update rate $\tau=0.01$, and maximum gradient clipping value $\xi=0.5$.
\begin{figure*}[t]
\centering
    \includegraphics[scale=0.6]{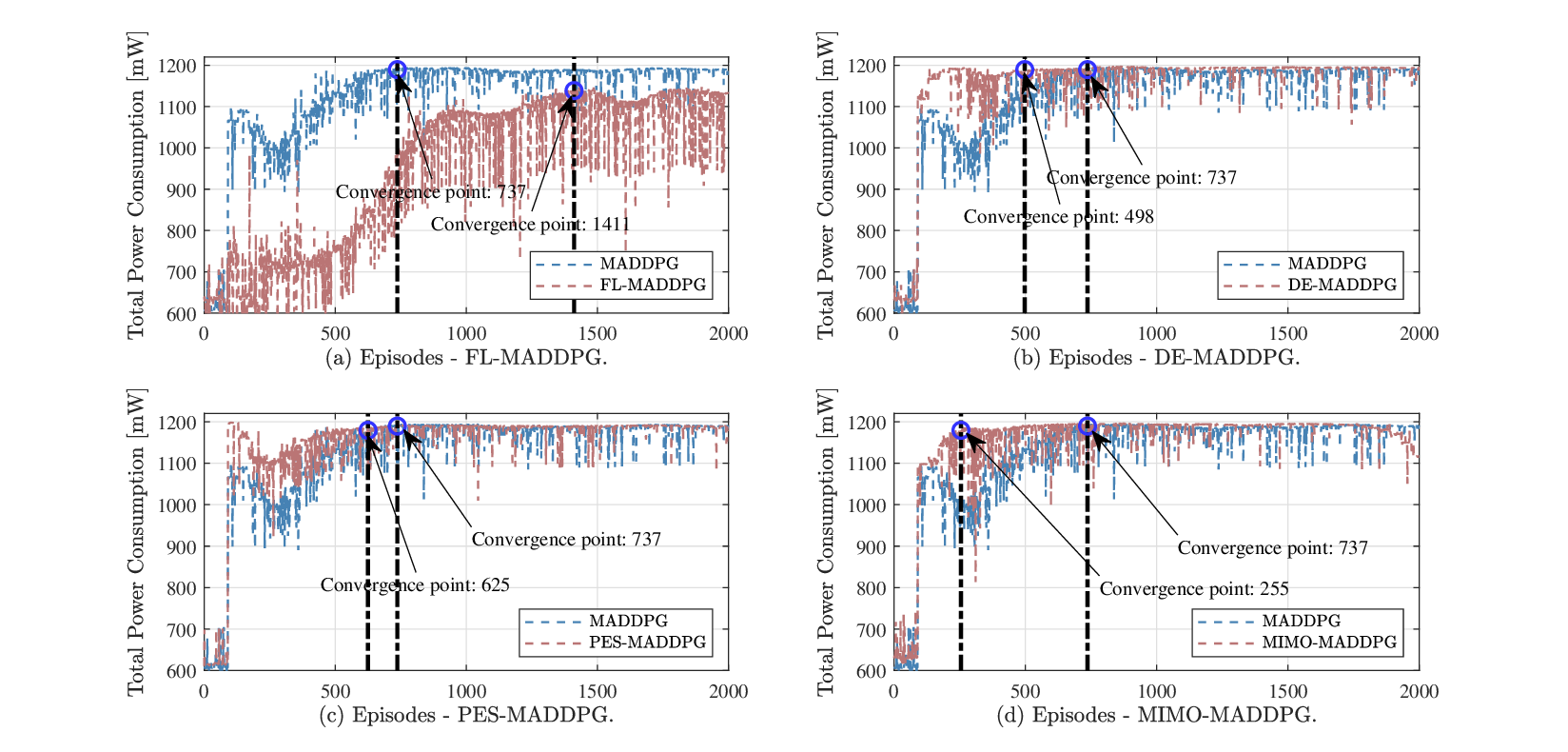}
    \caption{Convergence rate over different algorithm architectures in static scenarios for MR combining.
    \label{fig44}}
\end{figure*}
\begin{figure*}[t]
\centering
    \includegraphics[scale=0.6]{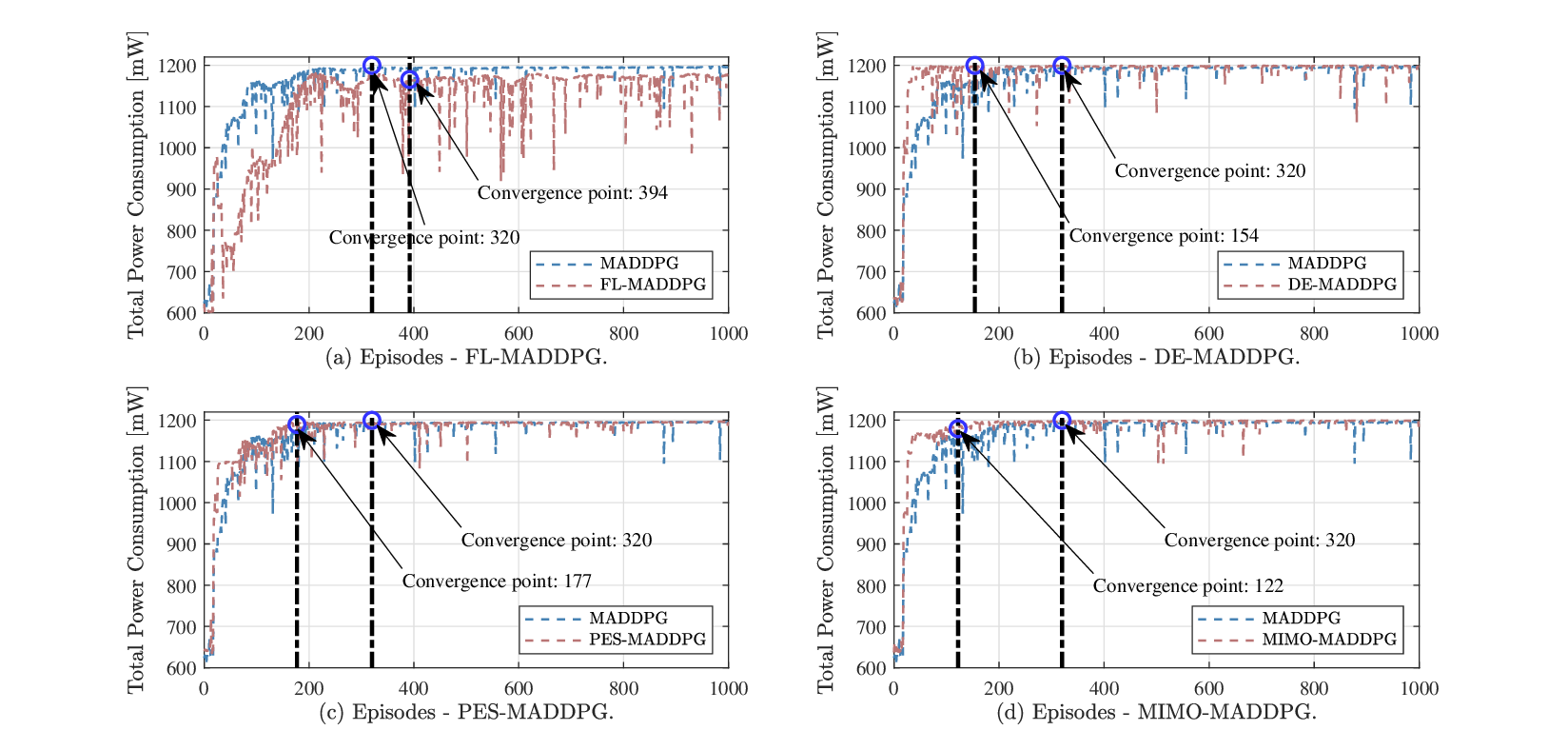}
    \caption{Convergence rate over different algorithm architectures in dynamic scenarios for MR combining.
    \label{fig55}}
\end{figure*}
\begin{table*}[t]
  \centering
  \fontsize{9}{12.15}\selectfont
  \caption{Comparison of convergence time for five MARL-based algorithms.}
  \label{Paper_comparison}
    \begin{tabular}{!{\vrule width1.2pt}  m{2.5 cm}<{\centering} !{\vrule width1.2pt}  m{3.4 cm}<{\centering} !{\vrule width1.2pt} m{1.1 cm}<{\centering} !{\vrule width1.2pt} m{1.1 cm}<{\centering} !{\vrule width1.2pt}  m{1.1 cm}<{\centering} !{\vrule width1.2pt} m{1.1 cm}<{\centering} !{\vrule width1.2pt} m{1.1 cm}<{\centering} !{\vrule width1.2pt}  m{1.1 cm}<{\centering} !{\vrule width1.2pt}}

    \Xhline{1.2pt}
            \multirow{2}{*}{\textbf{Methods}}  & \multirow{2}{*}{\textbf{Average training time [s]}} &  \multicolumn{2}{c!{\vrule width1.2pt}}{\textbf{Convergence point}}   & \multicolumn{2}{c!{\vrule width1.2pt}}{\textbf{Convergence [s]}}  & \multicolumn{2}{c!{\vrule width1.2pt}}{\textbf{Improvement [\%]}}  \\            \Xcline{3-8}{1.2pt}
          \multicolumn{1}{!{\vrule width1.2pt}c!{\vrule width1.2pt}}{\textbf{}} & \multicolumn{1}{c!{\vrule width1.2pt}}{} & \multicolumn{1}{c!{\vrule width1.2pt}}{Static} & \multicolumn{1}{c!{\vrule width1.2pt}}{Dynamic} & \multicolumn{1}{c!{\vrule width1.2pt}}{Static} & \multicolumn{1}{c!{\vrule width1.2pt}}{Dynamic} & \multicolumn{1}{c!{\vrule width1.2pt}}{Static} & \multicolumn{1}{c!{\vrule width1.2pt}}{Dynamic}  \\
    \Xhline{1.2pt}
        \cellcolor{gray!30} \bf MADDPG & 0.556 & 737 & 320 & 409 & 178 & --- & --- \cr\hline
        \cellcolor{gray!30} \bf FL-MADDPG & 0.431 & 1411 & 394 & 707 & 171 &  $\searrow$   & 4.5 \cr\hline
        \cellcolor{gray!30} \bf DE-MADDPG & 0.591 & 498 & 154 &  294 & 91 & 28.2 & 48.9 \cr\hline
        \cellcolor{gray!30} \bf PES-MADDPG & 0.567 & 625 & 177 & 354 & 101 & 13.5 & 43.6 \cr\hline
        \cellcolor{gray!30} \bf MIMO-MADDPG & 0.614 & 255 & 122 & 157 & 75 & 61.7 & 57.8 \cr\hline
    \Xhline{1.2pt}
    \end{tabular}
  \vspace{0cm}
\end{table*}
\subsection{Effects of the proposed MIMO-MADDPG method}
In this subsection, we first investigate the effects of different types of state information on system performance in the MARL environment, including LSF information and channel state information. As shown in Fig. 6, as the number of antennas at BSs increases, the SE performance of the LSF-based MADDPG scheme is approximately close to that of the unified channel-based MADDPG scheme. This shows that in the mobile scenario studied, the normalized instantaneous channel gain converges to the deterministic average channel gain as the number of antennas increases. Furthermore, since only large-scale fading is used as the observable state in the MARL environment, it helps to reduce the interaction information of agents and the computational complexity of the network.

Moreover, we investigate the effects of the designed MIMO-MADDPG and double-layer architecture. Fig. 7 illustrates the achievable sum SE over single-layer architecture and double-layer architecture with different MADDPG-based algorithms, such as FL-MADDPG in \cite{[2]}, DE-MADDPG in \cite{[6]}, PES-MADDPG in \cite{[7]}, and the proposed MIMO-MADDPG. We notice that the sum SE performance of the proposed MIMO-MADDPG algorithm is closer to the original MADDPG algorithm, which implies that the combination of the global critic network and prioritized experience selected mechanism is conducive to the improvement of SE performance. Moreover, the introduction of double-layer architecture has significantly improved the system performance, by enabling reasonable power allocation to each antenna for reducing interference between antennas. Compared with the single-layer architecture, the double-layer architecture has achieved 11.64\%, 11.72\% 11.67\%, and 11.75\% performance improvements for FL-MADDPG, DE-MADDPG, PES-MADDPG, and the proposed MIMO-MADDPG, respectively.

Fig. 8 shows the achievable sum SE over different MARL-based algorithms against the number of UEs and BSs. Numerical results show that our proposed MIMO-MADDPG algorithm scales well as the number of UEs and BSs increases, and always maintains the SE performance close to the conventional MADDPG algorithm. Moreover, Fig. 9 shows the achievable sum SE over two system architectures against the number of antennas per row at UEs. The SE performance gap between double-layer architecture and the single-layer architecture under two MADDPG-based algorithms increases with the number of antennas per row at UEs, e.g., the performance gaps are 11.77\% and 23.78\% over $N_{H_s}=N_{V_s}=2$ and $N_{H_s}=N_{V_s}=7$, respectively.
In Fig. 10, the achievable sum SE over two system architectures against the number of antennas per row at BSs is plotted. Compared with Fig. 9, although the SE performance in double-layer architecture is significantly better than that in single-layer architecture, the SE performance gap between the two architectures is almost independent of the number of antennas at BSs and fluctuates around 16\%.

To further demonstrate the advantage of the proposed double-layer architecture, Fig. 11 and Fig. 12 investigate the effects of antenna spacing at UEs and BSs on the sum SE performance improvement under the double-layer architecture, respectively. The figures reveal that the SE performance gap between the double-layer architecture and the single-layer architecture increases as the antenna spacing decreases. For example, compared with the single-layer architecture, the double-layer architecture yields a 20.89\% and 21.27\% improvement in sum SE under smaller antenna spacing.
\subsection{Comparison of Convergence Rate}
In this subsection, we investigate the convergence rate of different MADDPG-based algorithms with MR combining. For the convergence definition of the training curve, we take the starting point when the final training value of the agent stays within the small tolerance range of $\delta_{conv}$ in the course of $N_{conv}$ successive iterations as the convergence point. Generally, in a particular mMIMO scenario, we can set the number of iterations $N_{conv}$ and the error tolerance $\delta_{conv}$ to $100$ and $\pm1$\%, respectively. Fig. 13 shows a comparison of the convergence rate in static scenarios with $M=9$, $K=6$, $N_r=N_{H_r}\times N_{V_r}=100$, $N_s=N_{H_s}\times N_{V_s}=16$, and $\Delta_r = \lambda/4$. For the proposed MIMO-MADDPG algorithm, we combine the global critical network in DE-MADDPG and the prioritized experience selected mechanism in PES-MADDPG to improve convergence stability improved and optimize convergence rate.
We observe that the proposed MIMO-MADDPG algorithm demonstrates a 65.41\% improvement in convergence rate compared with the common MADDPG algorithm. Additionally, it outperforms the DE-MADDPG and PES-MADDPG algorithms with improvement effects of 32.43\% and 15.19\%, respectively.

Fig. 14 presents a comparison of the convergence rate in dynamic scenarios with $M=9$, $K=6$, $N_r=N_{H_r}\times N_{V_r}=100$, $N_s=N_{H_s}\times N_{V_s}=16$, and $\Delta_r = \lambda/4$.
Similar to the results in Fig. 13, the proposed MIMO-MADDPG algorithm shows a faster convergence rate in dynamic scenarios due to the combination of multiple optimization mechanisms. Specifically, the proposed algorithm demonstrates a 61.88\% improvement in convergence rate compared to the conventional MADDPG algorithms. In addition, considering the superior performance of MIMO-MADDPG in convergence speed analyzed above, the ability of real-time interaction can be further improved by combining new optimization tools including parallel computing, model compression, and transfer learning, etc., to meet the strict requirement of real-time and highly reliable communications to a certain extent.

Moreover, Table \uppercase\expandafter{\romannumeral4} presents a comparison of the convergence time with different MADDPG-based algorithms.
Despite the high training time or computational complexity per episode, the proposed MIMO-MADDPG algorithm significantly reduces the convergence time of the algorithm due to its faster convergence rate. Compared with the conventional MADDPG algorithm, the convergence time of the MIMO-MADDPG algorithm in static and dynamic scenarios has been reduced by 61.74\% and 57.84\%, respectively.
\section{Conclusion}
In this paper, we investigated the uplink SE performance of a CF XL-MIMO system over the near-field communication domain, where both the BSs and UEs are equipped with XL-MIMO panels. We considered a novel MARL-based paradigm for large-scale mMIMO systems, which combines the decoupling architecture in DE-MADDPG, the priority experience selected mechanism in PES-MADDPG, and predictive management for dynamic scenarios.
Furthermore, we introduced a double-layer power control architecture based on the unique near-field characteristics between antennas.
The numerical results showed that the proposed MIMO-MADDPG algorithm has a faster convergence rate and similar SE performance compared with the conventional MADDPG algorithms.
More significant, the SE performance of the proposed double-layer architecture was superior to that of the single-layer architecture, due to the double-layer architecture plays a significant role in reducing interference between antennas.
In future work, we will investigate uplink power control with hybrid channel estimation using the proposed MIMO-MADDPG algorithm, as well as the significant CF XL-MIMO channel characteristics.
\bibliographystyle{IEEEtran}
\bibliography{IEEEabrv,Ref}
\end{document}